\documentclass[review]{elsarticle}
\usepackage{amsfonts}
\usepackage{amsmath}
\usepackage{amssymb}
\usepackage{graphicx}
\usepackage{xcolor}
\usepackage{setspace}
\usepackage{CJK}
\usepackage{moresize}
\usepackage[english]{babel}
\usepackage{lineno,hyperref}

\def\){\Big)}
\def\({\Big(}
\def\a{a}
\def\b{b}
\def\c{c}
\def\d{d}

\begin{document}

\title{Noether and Hilbert (metric) energy-momentum tensors are not, in general, equivalent}
\author[mymainaddress,mysecondaryaddress]{Mark Robert Baker}
\ead{mbaker66@uwo.ca}
\corref{mycorrespondingauthor}
\cortext[mycorrespondingauthor]{Corresponding author}
\author[mythirdaddress]{Natalia Kiriushcheva}
\ead{nkiriush@uwo.ca}
\author[mymainaddress,myfourthaddress]{Sergei Kuzmin}
\ead{skuzmin@uwo.ca}

\address[mymainaddress]{Department of Physics and Astronomy, Western University, London, ON, Canada N6A 5B7}
\address[mysecondaryaddress]{The Rotman Institute of Philosophy, Western University, London, ON, Canada N6A 5B7}
\address[mythirdaddress]{Department of Applied Mathematics, Western University,  London, ON, Canada N6A 5B7}
\address[myfourthaddress]{School of Management, Economics, and Mathematics, King's University College at Western University, London, ON, Canada N6A 2M3}

\date{\today}

\begin{abstract}
Multiple methods for deriving the energy-momentum tensor for a physical
theory exist in the literature. The most common methods are to use Noether's
first theorem with the 4-parameter Poincar\'{e} translation, or to write
the action in a curved spacetime and perform variation with respect to
the metric tensor, then return to a Minkowski spacetime. These are referred
to as the Noether and Hilbert (metric/ curved space/ variational) energy-momentum
tensors, respectively. In electrodynamics and other simple models, the Noether and Hilbert methods yield the same result. Due to this fact, it is often asserted
that these methods are generally equivalent for any theory considered,
and that this gives physicists a freedom in using either method to derive
an energy-momentum tensor depending on the problem at hand. The ambiguity
in selecting one of these two different methods has gained attention in
the literature, but the best attempted proofs of equivalence of
the two methods require restrictions on the order of derivatives and rank of fields; general equivalence of the Noether and Hilbert methods has not been established. For spin-2, the ideal candidate to check this equivalence
for a more complicated model, there exist many energy-momentum tensors
in the literature, none of which are gauge invariant, so it is not clear
which expression one hopes to obtain from the Noether and Hilbert approaches
unlike in the case of e.g. electrodynamics. It has been shown, however,
that the linearized Gauss-Bonnet gravity model (second order derivatives,
second rank tensor potential) has an energy-momentum tensor that is unique,
gauge invariant, symmetric, conserved, and trace-free when derived from
Noether's first theorem (all the same properties of the physical energy-momentum
tensor of electrodynamics). This makes it the ideal candidate to check
if the Noether and Hilbert methods coincide for a more complicated model.
It is proven here using this model as a counterexample, by direct calculation,
that the Noether and Hilbert energy-momentum tensors are not, in general,
equivalent.
\end{abstract}

\maketitle

\linenumbers

\nolinenumbers

\section{Motivation}

The energy-momentum tensor of a physical field theory is an expression of fundamental significance to a physical model. In electrodynamics, for example, it compactly expresses familiar conservation laws and the Lorentz force law upon differentiation. However, the procedure to derive an energy-momentum tensor from a Lagrangian density is not unique. Several methods for deriving this expression can be found in the literature. We will focus on the two most common procedures for deriving an energy-momentum tensor in Minkowski spacetime, the Noether and Hilbert methods \cite{blaschke2016}. Methods such as the Fock method \cite{fock2015,bicak2016} will not be discussed in this article, as they don't involve a procedure to derive the energy-momentum tensor from a Lagrangian density.

For a recent summary of the Noether and Hilbert methods, we will refer the reader to the paper by Blaschke, Gieres, Reboud and Schweda `The energy-momentum tensor(s) in classical gauge theories' published in Nuclear Physics B in 2016 \cite{blaschke2016}. We will refer to the paper as BGRS \cite{blaschke2016} due to the frequent reference to their paper our article. The BGRS paper has an extensive summary of the literature, so we suggest turning to BGRS \cite{blaschke2016} and the references therein if the reader is unfamiliar with these topics. It is well known that for a vector field (electrodynamics) and for a scalar field (Klein-Gordon), the Noether and Hilbert methods coincide with the same energy-momentum tensor. In BGRS \cite{blaschke2016} the authors address this question for Yang-Mills and spinor fields and conclude again that these are equivalent from both Noether and Hilbert approaches, yet again those are models with at most first order derivatives of a vector potential in the action. The best attempted proof of
equivalence of the Noether and Hilbert methods has been limited to simple
models with at most first order derivatives in the
action \cite{borokhov2002,saravi2004,forger2004,leclerc2006,pons2011}.
Unfortunately none of these authors considered a more complicated model
to test the equivalence of the Noether and Hilbert methods in Minkowski spacetime, as it only takes one counterexample to disprove
the notion of general equivalence; this is what will be provided in the
present article.

In this article we will focus on the very specific question: for more complicated models in Minkowski spacetime, do the Noether and Hilbert methods yield an equivalent result? In other words, do actions with higher order derivatives and higher ranks of tensor potential, such as the linearized higher derivative gravity models, yield the same energy-momentum tensor by following the Noether and Hilbert procedures. The ideal candidate to explore this question, spin-2, is problematic because it has been proven that there exists no gauge invariant energy-momentum tensor for that model \cite{magnano2002}, and there is no generally accepted unique energy-momentum tensor for the theory, as many exist in the literature \cite{padmanabhan2008,bicak2016}. This issue has come to the forefront recently regarding the necessity to have a well defined energy-momentum tensor for the spin-2 field to self couple in the standard spin-2 to general relativity derivations \cite{padmanabhan2008,butcher2009,deser2010,bicak2016}. Therefore it is not clear which expression one hopes to obtain from both the Noether and Hilbert method for spin-2 as in the case of electrodynamics where a single, accepted physical tensor exists.

A more complicated ideal candidate does exist, in the form of the linearized higher derivative gravity models, that is the models built from the contracted linearized Riemann tensor $R_{\mu\nu\alpha\beta} R^{\mu\nu\alpha\beta}$, Ricci tensor $R_{\mu\nu} R^{\mu\nu}$ and Ricci scalar $R^2$. These relativistic models in Minkowski spacetime have second order derivatives of a second rank symmetric tensor potential $h_{\mu\nu}$ in the action (terms of the form $\partial \partial h \partial \partial h$). In particular, we will consider the energy-momentum tensor for the linearized Gauss-Bonnet gravity model, which has been well known to string theorists and other researchers for some time \cite{myers1987}. This expression has been shown to be derived from Noether's first theorem \cite{baker2019}, and it is unique, gauge invariant, symmetric, conserved, and trace-free; all the properties of a physical energy-momentum tensor as defined by BGRS \cite{blaschke2016}. It is these properties of electrodynamics that allows the equivalence between the Noether and Hilbert methods to be accurately concluded. Using this example of the linearized higher derivative gravity models, and in particular the linearized Gauss-Bonnet gravity model, we give a proof by counterexample that the Noether and Hilbert energy-momentum tensors are not, in general, equivalent. We then outline why this result will hold for higher order of derivative/ higher rank of tensor potential models more generally. If several methods exist, and they do not generally yield the same result, it is an issue of fundamental significance as to which method is truly allowing one to derive physical results for any general action, and which happen to coincide for actions of simple physical models.

\section{The Noether and Hilbert methods for deriving energy-momentum tensors in Minkowski spacetime}

From Noether's theorem the energy-momentum tensor for electrodynamics was directly derived by Bessel-Hagen in 1921 \cite{besselhagen1921}, without the need for improvements, by considering the gauge symmetry of the action. Several other authors came to a similar conclusion later \cite{burgess2002, montesinos2006, eriksen1980,munoz1996}, apparently unaware of Bessel-Hagen's paper, which was only recently translated into English \cite{besseltranslation(2006)}. If the action is exactly gauge invariant, this procedure derives the physical energy-momentum tensor for the theory without the need to add any ad-hoc improvement terms to obtain a gauge invariant expression. In BGRS \cite{blaschke2016}, the authors outline this procedure in section 2.2.2, but without referring to Bessel-Hagen, only to \cite{montesinos2006, eriksen1980,munoz1996}. We will refer to this as the Bessel-Hagen method because he was the first to present this procedure, and in our opinion, in the clearest and most direct way based on Noether's original work.

It is important to briefly mention the `improvement' of energy-momentum tensors derived from Noether's theorem in the literature, due to its widespread use. Various improvements exist and are well summarized in BGRS \cite{blaschke2016}. Conventional wisdom states that one can improve Noether's energy-momentum tensor when the result one obtains from Noether's first theorem is not the physically accepted expression for the energy-momentum tensor. This involves adding terms which do not follow from Noether's theorem in order to obtain the desired result. Since the Bessel-Hagen method derives the correct, physical expressions directly from Noether's theorem without the need to add any terms, it is not really an `improvement' (no ad-hoc terms need to be added), rather it is the correct derivation intended by Noether, who is cited as giving Bessel-Hagen the ideas for his paper. We note that the most common improvement found in the literature is the Belinfante method \cite{belinfante1940}, designed to build a symmetric energy-momentum tensor from the non-physical `canonical Noether' energy-momentum tensor. This improvement does not guarantee gauge invariance of the energy-momentum tensor, a deficiency addressed by a new improvement procedure of BGRS \cite{blaschke2016}.

This sentiment was summarized in BGRS \cite{blaschke2016}, namely the importance of a gauge invariant energy-momentum tensor for theories considered physical, and the deficiencies of the Belinfante method \cite{belinfante1940}: {\it{`If one considers gauge field theories in Minkowski space as we do in the present article, then the EMT}} [energy-momentum tensor] {\it{necessarily has to be gauge invariant due to its physical interpretation. However, Belinfante’s improvement procedure does not yield a priori a gauge invariant EMT  when applied to gauge theories, and in addition it does not work in the straightforward manner for the physically interesting case where matter fields are minimally coupled to a gauge field.'}}  In the cases of electrodynamics and linearized Gauss-Bonnet gravity, the accepted physical, unique, gauge invariant, symmetric, conserved, and trace-free expressions are obtained from the Bessel-Hagen method, so there is no need to add improvement terms to the Noether result for these models.

Noether's first theorem is used to derive conservation laws by considering the action $S=\int \mathcal{L} dx$ to be invariant under simultaneous variation of the coordinates $\delta x_\nu$ and fields $\delta \Phi_A$ (where $\mathcal{L}(\Phi_A, \partial_\mu \Phi_A, \partial_\mu \partial_\nu \Phi_A, \dots)$ is the Lagrangian density, $A$ represents any rank of tensor potential $\Phi_A$ and $\partial_\mu=\frac{\partial}{\partial x^\mu}$ is abbreviated notation for a derivative). From Noether's first theorem we have the relationship between the Euler-Lagrange equation and some total derivative  \cite{noether1918,kosmann2011noether,gelfand2012},

\begin{multline}
\left( \frac{\partial \mathcal{L}}{\partial \Phi_A}
 - \partial_\mu \frac{\partial \mathcal{L}}{\partial (\partial_\mu \Phi_A)} 
 + \partial_\mu \partial_\omega \frac{\partial \mathcal{L}}{\partial (\partial_\mu \partial_\omega \Phi_A)} + \dots \ \right) \delta \Phi_A
 \\
+ \partial_\mu \left(  \eta^{\mu\nu} \mathcal{L} \delta x_\nu
+ \frac{\partial \mathcal{L}}{\partial (\partial_\mu \Phi_A)} \delta \Phi_A 
+ \frac{\partial \mathcal{L}}{\partial (\partial_\mu \partial_\omega \Phi_A)} \partial_\omega \delta \Phi_A
- \left[ \partial_\omega \frac{\partial \mathcal{L}}{\partial (\partial_\mu \partial_\omega \Phi_A)} \right] \delta \Phi_A
+ ... \ \right) = 0 \ . 
   \label{genenergy}
\end{multline}

Using Equation (\ref{genenergy}) and the Bessel-Hagen method we can derive the standard energy-momentum tensor $T^{\mu\nu}_N = F^{\mu\alpha} F^\nu_{\ \alpha} - \frac{1}{4} \eta^{\mu\nu} F_{\alpha\beta} F^{\alpha\beta}$ for electrodynamics (with the field strength $F_{\alpha \beta} =\partial_\alpha A_\beta - \partial_\beta A_\alpha$) from Noether's theorem by use of the 4-parameter Poincar{\'e} translation and gauge invariance of the action, where subscript $N$ will be used to identify any physical expression derived from Noether's theorem.

The other most common procedure for deriving an energy-momentum tensor in Minkowski spacetime is the Hilbert method, sometimes referred to as the metric energy-momentum tensor, curved space energy-momentum tensor, and even variational energy-momentum tensor. A good summary of this method is found in BGRS \cite{blaschke2016}, Section 3 `Einstein-Hilbert EMT in Minkowski space'. The authors refer to this tensor as the metric energy-momentum tensor in their article. The Hilbert energy-momentum tensor is derived by writing the action in `curved space' by replacing all ordinary derivatives with covariant derivatives $\partial \to \nabla$, replacing the Minkowski metric with the general metric tensor $\eta \to g$, and inserting the Jacobian term $\sqrt{-g}$. After expressing the action in this form, the variation with respect to the general metric tensor is performed,

\begin{equation}
\frac{\delta \mathcal{L}}{\delta  g_{\gamma\rho}} =  \frac{\partial \mathcal{L}}{\partial g_{\gamma\rho}} 
- \partial_\omega \frac{\partial \mathcal{L}}{\partial (\partial_\omega g_{\gamma\rho})}
+  \partial_\xi \partial_\omega \frac{\partial \mathcal{L}}{\partial (\partial_\xi \partial_\omega g_{\gamma\rho})}
+ \dots \quad .
\end{equation}

Once the variational derivative is found from this procedure, it is then `returned to flat space' by replacing the metric tensors with the Minkowski metric, yielding an energy-momentum tensor of the form,

\begin{equation}
T^{\gamma\rho}_H = \frac{2}{\sqrt{-g}} \frac{\delta \mathcal{L}}{\delta  g_{\gamma\rho}} \Bigg|_{g = \eta} . \label{genhilbert}
\end{equation}

Note that this definition is given in equation (3.18) in BGRS \cite{blaschke2016}, where we take the $+$ instead of $-$ expression here so that the signs match the derivation for electrodynamics in the following section (both signs can be found throughout the literature depending on convention). The subscript $H$ will indicate what is derived from the Hilbert method. Remarkably, for electrodynamics, these two expressions coincide $T^{\gamma\rho}_N = T^{\gamma\rho}_H$. Due to this coincidence, and the fact that electrodynamics has a unique physical energy-momentum tensor accepted in the literature, it is tempting to assert general statements about their equivalence. Other simple models amplify these sentiments, leading to the belief that the results of these methods are in some sense generally equivalent. However, they have only been reconciled for simple scalar or vector fields and first order derivatives in actions. Higher order of derivative, higher rank of tensor potential models, such as those presented in this article, have not previously been considered to verify the general equivalence of the Noether and Hilbert methods.

In BGRS \cite{blaschke2016}, the authors remark about the Noether tensor $T^{\gamma\rho}_N$ (including improvements) vs. the Hilbert tensor $T^{\gamma\rho}_H$ by stating {\it{`This definition of the EMT in Minkowski space is conceptually and mathematically quite different from the one of $T^{\mu\nu}_{imp} [\psi]$ which we presented in section 2 and which follows from Noether’s theorem (eventually supplemented by an improvement procedure to render the canonical expression of the EMT symmetric in its indices or gauge invariant, or both symmetric and traceless).'}} They go on to consider at most first order, vector models as is common in the literature {\it{`In the following, we will show that the two definitions for the EMT’s of YM-theories in Minkowski space,}} [...]{\it {which results from the coupling to gravity, and the improved EMT}} [...]{\it {which follows from Noether’s first theorem supplemented by the “gauge improvement” procedure, coincide with each other.'}}. We note that the Bessel-Hagen method can be used to derive the physical energy-momentum tensor for electrodynamics and Yang-Mills theory directly from Noether's theorem without the need for any improvements. This is why a higher derivative model, such as a linearized higher derivative gravity model, is so important to consider. To explore the question of general equivalence between the Noether and Hilbert methods, we must check if they coincide beyond simple physical models that we already know.

\section{Equivalence of the Noether and Hilbert expressions for classical electrodynamics}

Before comparing the Noether $T^{\gamma\rho}_N$ and Hilbert $T^{\gamma\rho}_H$ for the linearized Gauss-Bonnet model, it is best to recap the equivalence $T^{\gamma\rho}_N = T^{\gamma\rho}_H$ for electrodynamics, to show how to perform these derivations for a simple model before moving on to the higher order case. The physical energy-momentum tensor for electrodynamics, which was known before the publication of Noether's theorems, was first derived by Bessel-Hagen  in 1921 \cite{besselhagen1921}. Using equation (\ref{genenergy}), he derived this expression directly from the standard electromagnetic Lagrangian density $\mathcal{L} = - \frac{1}{4} F_{\alpha\beta} F^{\alpha\beta}$ for the field strength tensor $F_{\alpha\beta}=\partial_\alpha A_\beta - \partial_\beta A_\alpha$. For a theory with a Lagrangian built from terms quadratic in first order derivatives of a vector potential, Equation (\ref{genenergy}) simplifies to,

\begin{equation}
\left(  \partial_\gamma  \frac{\partial \mathcal{L}}{\partial (\partial_\gamma A_\nu)} \right)  \delta A_\nu = \partial_\gamma \left(  \eta^{\gamma\nu} \mathcal{L} \delta x_\nu
+ \frac{\partial \mathcal{L}}{\partial (\partial_\gamma A_\nu)} \delta A_\nu
 \right)  , \label{eq4}
\end{equation}

where the equation of motion (left hand side) forms an identity with the
conservation law (right hand side). For a conformally invariant theory
such as electrodynamics, the transformations of the 15 parameter conformal group that will leave the action invariant is $\delta x_\alpha = a_\alpha + \omega_{\alpha\beta} x^\beta + S x_\alpha + 2 \xi_\nu x_\alpha x^\nu  - \xi_\alpha x_\nu x^\nu$. The first term, the 4 parameter translation of the Poincar{\'e} group, is the `symmetry' that corresponds to energy-momentum tensors derived from Noether's theorem. The transformation of fields $\delta A_\nu$ that leave the action invariant are defined generally for Noether's first theorem as $\delta A_\nu = \delta A'_\nu - \partial^\beta A_\nu \delta x_\beta $ \cite{noether1918,kosmann2011noether,besselhagen1921,gelfand2012}. The first term, $\delta A'_\nu$, is related to field transformations that leave the action invariant (ie the spin-1 gauge transformation); this was neither discussed nor specified by Noether and could be anything (i.e., gauge symmetries) that preserves invariance of the action. Bessel-Hagen showed that using gauge invariance of the action to define $\delta A'_\nu$ that the transformation of the potential is exactly $\delta A_\nu = F_{\nu \rho} \delta x^\rho$. Inserting this $\delta A_{\nu }$, as well as
$\delta x_{\rho }= a_{\rho }$ and
$\frac{\partial \mathcal{L}}{\partial (\partial _{\gamma }A_{\nu })} = - F^{
\gamma \nu }$ into Equation (\ref{eq4}) we have,

\begin{equation}
\left(  - \partial_\gamma F^{\gamma\nu} \right)  \delta A_\nu = a_\rho\partial_\gamma \left(F^{\gamma\nu} F^\rho_{\ \nu}  -  \frac{1}{4} \eta^{\gamma\rho} F_{\alpha\beta} F^{\alpha\beta}
 \right) .
\end{equation}

Therefore the energy-momentum tensor for electrodynamic theory $T^{\gamma\rho}_N = F^{\gamma\nu} F^\rho_{\ \nu}  -  \frac{1}{4} \eta^{\gamma\rho} F_{\alpha\beta} F^{\alpha\beta}$ is derived directly from Noether's first theorem. 

The Hilbert energy-momentum tensor for electrodynamics is derived by equation (\ref{genhilbert}) after expressing the standard Lagrangian in curved space form, namely replacing the Minkowski metrics with the metric tensor, replacing all ordinary derivatives with covariant derivatives $\partial \to \nabla$, replacing all Minkowski metrics with general metrics $\eta \to g$, and inserting the Jacobian  $\sqrt{-g}$. Starting by re-writing the field strength tensor in terms of covariant derivatives $F^\nabla_{\mu\nu} = \nabla_\mu A_\nu - \nabla_\nu A_\mu = \partial_\mu A_\nu - \Gamma^\alpha_{\mu\nu} A_\alpha - \partial_\nu A_\mu + \Gamma^\alpha_{\nu\mu} A_\alpha = F_{\mu\nu}$, we see that the $\Gamma$ part exactly cancels itself, recovering the original field strength tensor. It should be noticed that for higher derivative models of higher rank potentials, many extra $\Gamma$ parts remain without cancellation, creating many more terms in the final energy-momentum tensor. This is likely part of the reason why for simple models the two methods $T^{\gamma\rho}_N = T^{\gamma\rho}_H$ coincide. Therefore for the curved space Lagrangian density we have,

\begin{equation}
\mathcal{L} = - \frac{1}{4} \sqrt{-g} g^{\alpha\mu} g^{\beta\nu} F_{\alpha\beta} F_{\mu\nu} .
\end{equation}

This simplifies the Euler derivative to just including derivatives of the metric, leaving for the Hilbert energy-momentum tensor $T^{\gamma\rho}_H = \frac{2}{\sqrt{-g}} \frac{\partial \mathcal{L} }{\partial  g_{\gamma\rho}} |_{g = \eta} $. Taking the derivative with respect to the metric, we use $ \frac{\partial \sqrt{-g}}{\partial g_{\gamma\rho}} =  \frac{1}{2} g^{\gamma\rho} \sqrt{-g}$ and $ \frac{\partial g^{\lambda\nu}}{ \partial g_{\beta\gamma}} = - \frac{1}{2} ( g^{\beta\lambda} g^{\gamma\nu} +  g^{\gamma\lambda} g^{\beta\nu})$. Performing this differentiation we have $\frac{\partial \mathcal{L}}{\partial  g_{\gamma\rho}}  = \frac{1}{2} \sqrt{-g} F_{\alpha\beta} F_{\mu\nu} (g^{\nu\beta} g^{\rho\mu} g^{\gamma\alpha} - \frac{1}{4} g^{\gamma\rho} g^{\beta\nu} g^{\alpha\mu} )$. Therefore $T^{\gamma\rho}_H = \frac{2}{\sqrt{-g}} \frac{\partial \mathcal{L}}{\partial  g_{\gamma\rho}} |_{g = \eta} = F^{\gamma\nu} F^\rho_{\ \nu}  -  \frac{1}{4} \eta^{\gamma\rho} F_{\alpha\beta} F^{\alpha\beta}$ which exactly coincides with what is derived from the Noether method, $T^{\gamma\rho}_N = T^{\gamma\rho}_H$.

The fact that the two derivations yield the same result is of fundamental interest since the two methods are mathematically quite different, as the authors of BGRS \cite{blaschke2016} noted. The problem is, these two expressions are only ever calculated for simple models with first order derivatives of at most a vector potential in the action. Both attempts at a general proof \cite{forger2004,pons2011} also rely on these simple models. We will now consider the linearized Gauss-Bonnet model which has a physical, unique, symmetric, gauge invariant, conserved and trace-free energy-momentum tensor derived using Noether's first theorem, as in the case of electrodynamics. As we will see, this greatly complicates the Hilbert expression due to second order derivatives and second rank tensor potential of the model, yielding a proof by counterexample that the Noether and Hilbert energy-momentum tensors are not generally equivalent.

\section{Non-Equivalence of the Noether and Hilbert expressions for Linearized Gauss-Bonnet gravity}

We will now consider the linearized Gauss-Bonnet gravity model (a relativistic model in Minkowski spacetime) that has a well known energy-momentum tensor derived from the Noether method. Here we will derive the Hilbert (metric) energy-momentum tensor in Minkowski spacetime as outlined by BGRS \cite{blaschke2016}, and compare to the Noether result to see if they are truly equivalent for this more complicated model. The Lagrangian density for this model is

\begin{equation}
\mathcal{L} = \frac{1}{4} (R_{\mu\nu\alpha\beta} R^{\mu\nu\alpha\beta} - 4 R_{\mu\nu} R^{\mu\nu} + R^2) ,
\end{equation}

where the scalars are built from contraction of the linearized Riemann tensor $R_{\mu\nu\alpha\beta}$, Ricci tensor $R_{\mu\nu}$ and Ricci scalar $R$:

\begin{gather}
R^{\mu\nu\alpha\beta} = \frac{1}{2} (  \partial^\mu \partial^\beta h^{\nu\alpha} + \partial^\nu \partial^\alpha h^{\mu\beta} -\partial^\mu \partial^\alpha h^{\nu\beta} - \partial^\nu \partial^\beta h^{\mu\alpha}) , \label{linR1}
\\
R^{\nu\beta} = \eta_{\mu\alpha} R^{\mu\nu\alpha\beta} = \frac{1}{2} (   \partial^\beta \partial^\alpha h_{\alpha}^{\nu}+\partial^\nu \partial^\alpha h_{\alpha}^{\beta} -\square h^{\nu\beta} - \partial^\nu \partial^\beta h ) , \label{linR2}
\\
R = \eta_{\nu\beta} R^{\nu\beta} =  \partial_\mu \partial_\nu h^{\mu\nu} - \square h . \label{linR3}
\end{gather}

Each of these $R$'s will indicate these linearized expressions unless otherwise noted. The energy-momentum tensor for Gauss-Bonnet gravity has been well known to string theorists and other researchers for some time \cite{myers1987},

\begin{equation}
T^{\omega\nu}_N = -  R^{\omega\rho\lambda\sigma} R^\nu_{\ \rho\lambda\sigma}
 + 2  R_{\rho\sigma} R^{\omega \rho \nu \sigma}
+ 2  R^{\omega\lambda} R^\nu_{\ \lambda}
 -  R R^{\omega\nu} + \frac{1}{4} \eta^{\omega\nu}(R_{\mu\lambda\alpha\beta} R^{\mu\lambda\alpha\beta} - 4 R_{\mu\gamma} R^{\mu\gamma} + R^2) . \label{gaussemt}
\end{equation}

This energy-momentum tensor is derived from Noether's first theorem, equation (\ref{genenergy}), for the linearized Gauss-Bonnet gravity model \cite{baker2019}. It is the unique, symmetric, gauge invariant, conserved and trace-free expression for the model, all properties of a physical energy momentum tensor as defined by BGRS \cite{blaschke2016}. This allows for an accurate comparison to be made between the Noether and Hilbert energy-momentum tensors, as in the case of electrodynamics. Since a uniquely defined Noether energy-momentum tensor can be derived from an action with second order derivatives and a second rank tensor potential, of the form $\partial \partial h \partial \partial h$, this model is the ideal candidate to test equivalency with the energy-momentum tensor derived from the Hilbert method in Minkowski spacetime. We will perform this derivation with free coefficients,

\begin{equation}
\mathcal{L} = A R_{\mu\nu\alpha\beta} R^{\mu\nu\alpha\beta} + B R_{\nu\beta} R^{\nu\beta} + C R^2 \label{genlagfree} ,  
\end{equation}

in case the reader is interested in the Hilbert energy-momentum tensor for other linearized modified gravity models. Expanding the Lagrangian in Equation (\ref{genlagfree}) in terms of Equations (\ref{linR1}), (\ref{linR2}) and (\ref{linR3}) we have,

\begin{multline}
\mathcal{L} = A (\partial_\mu \partial_\nu h_{\alpha\beta} \partial^\mu \partial^\nu h^{\alpha\beta} - 2 \partial_\mu \partial_\nu h_{\alpha\beta} \partial^\mu \partial^\alpha h^{\nu\beta} + \partial_\mu \partial_\nu h_{\alpha\beta} \partial^\alpha \partial^\beta h^{\mu\nu})
\\
+ \frac{1}{4} B ( \partial_\mu \partial^\mu h_{\alpha\beta} \partial_\nu \partial^\nu h^{\alpha\beta} + 2 \partial_\mu \partial_\nu h_{\alpha}^{\alpha} \partial_\beta \partial^\beta h^{\mu\nu} 
 - 4 \partial_\mu \partial_\nu h^{\nu}_{\beta} \partial_\alpha \partial^\alpha h^{\mu\beta} 
 \\
 + \partial_\mu \partial_\nu h_{\alpha}^{\alpha} \partial^\mu \partial^\nu h_{\beta}^{\beta} 
 - 4 \partial_\mu \partial_\nu h_{\alpha}^{\alpha} \partial^\mu \partial_\beta h^{\nu\beta} + 2 \partial_\mu \partial_\nu h^{\nu\beta} \partial^\mu \partial_\alpha h^\alpha_\beta + 2 \partial_\mu \partial_\nu h^\nu_{\beta} \partial^\beta \partial_\alpha h^{\mu\alpha} )
\\
+ C (\partial_\mu \partial^\mu h_\nu^{\nu} \partial_\alpha \partial^\alpha h_{\beta}^{\beta} - 2 \partial_\mu \partial_\nu h^{\mu\nu} \partial_\alpha \partial^\alpha h_{\beta}^{\beta} + \partial_\mu \partial_\nu h^{\mu\nu} \partial_\alpha \partial_\beta h^{\alpha\beta}) \label{expandedLGBL} .
\end{multline}

This is the expanded form of the Lagrangian density which we are using to compare the Noether and Hilbert methods. In other words, we are considering a relativistic model in Minkowski spacetime with terms $\partial \partial h \partial \partial h$ in the Lagrangian density. The Hilbert (metric) energy-momentum tensor in Minkowski spacetime has been considered for many models before, for example the spin-2 Fierz-Pauli Lagrangian density \cite{fierz1939} (see also \cite{bicak2016,padmanabhan2008}), $\mathcal{L}_{FP} =  \frac{1}{4} [\partial_\alpha h_\beta^\beta \partial^\alpha h_\gamma^\gamma 
- \partial_\alpha h_{\beta\gamma} \partial^\alpha h^{\beta\gamma} 
+ 2 \partial_\alpha h_{\beta\gamma} \partial^\gamma h^{\beta\alpha}
- 2 \partial^\alpha h_{\beta}^\beta \partial^\gamma h_{\gamma\alpha}] $. In \cite{fierz1939} Fierz and Pauli developed this action without reference to general metric spacetimes, it is a purely relativistic field theory in Minkowski spacetime. The spin-2 Hilbert energy-momentum tensor was calculated by \cite{bicak2016} in their Equation 30. It should be emphasized that $h_{\mu\nu}$ is a symmetric second rank tensor field of a special relativistic (Poincar{\'e} invariant) field theory in Minkowski spacetime; these $h_{\mu\nu}$ have no explicit or implicit dependence on the metric $g_{\mu\nu}$. The same goes for the linearized Gauss-Bonnet gravity model.  For the purpose of our disproof, this is just a relativistic model in Minkowski spacetime with derivatives of a second rank symmetric tensor potential $h_{\mu\nu}$ in the action (the $\partial \partial h \partial \partial h$ in Equation (\ref{expandedLGBL})). We use this model because it is sufficiently nontrivial to show that applying both the Noether and Hilbert methods to a common Lagrangian density can yield different results. These results hold more generally for other such nontrival models (higher order derivatives, higher rank of tensor potential), as outlined by the Reasons 1-3 in Section 5. We will now calculate the Hilbert (metric) energy-momentum tensor for this model.

\subsection{Expressing the Lagrangian in terms of the metric and covariant derivatives}

In order to derive the Hilbert energy-momentum tensor, we must replace in Equation (\ref{expandedLGBL}) all ordinary derivatives with covariant derivatives $\partial \to \nabla$, replacing all Minkowski metrics with general metrics $\eta \to g$, and inserting the Jacobian term $\sqrt{-g}$ to this action. In order for brevity, we will write Equation (\ref{expandedLGBL}) compactly as Equation (\ref{genlagfree}), thus the Lagrangian takes the form,

\begin{equation}
\mathcal{L} =  A \sqrt{-g} g^{a \mu} g^{b \nu} g^{c \alpha} g^{d \beta} R^\nabla_{abcd} R^\nabla_{\mu\nu\alpha\beta} + B  \sqrt{-g}  g^{\nu b} g^{\beta d} R^\nabla_{\nu\beta} R^\nabla_{bd} + C \sqrt{-g} R^\nabla R^\nabla ,  \label{alllingravlags}
\end{equation}

where a superscript $\nabla$ indicates that in the expressions from Equations (\ref{linR1}), (\ref{linR2}) and (\ref{linR3}), the linearized $R^{\mu\nu\alpha\beta}$, $R^{\nu\beta}$ and $R$ have their ordinary derivatives are replaced by covariant derivatives,

\begin{equation}
R^\nabla_{\mu\nu\alpha\beta} = \frac{1}{2} ( \nabla_\mu \nabla_\alpha h_{\nu\beta} + \nabla_\nu \nabla_\beta h_{\mu\alpha} - \nabla_\mu \nabla_\beta h_{\nu\alpha} - \nabla_\nu \nabla_\alpha h_{\mu\beta}) . \label{riecovdiv}
\end{equation}

Note that both Latin and Greek indices represent 4 dimensions ($a,b, \dots=1,2,3,4$ and $\alpha, \beta, \dots =1,2,3,4$). For the Ricci tensor, since it is defined in terms of the Riemann tensor $R_{\nu\beta} = \eta^{\mu\alpha} R_{\mu\nu\alpha\beta} $, we can express the covariant form in terms of the covariant Riemann tensor $R^\nabla_{\nu\beta} = g^{\mu\alpha} R^\nabla_{\mu\nu\alpha\beta}$. Similarly, the Ricci scalar can be expressed as $R^\nabla = g^{\nu\beta} g^{\mu\alpha} R^\nabla_{\mu\nu\alpha\beta}$. This allows the Lagrangian to be expressed entirely in terms of the metric and $R^\nabla_{\mu\nu\alpha\beta}$ of Equation (\ref{riecovdiv}),

\begin{equation}
\mathcal{L} =  \sqrt{-g} (A g^{a \mu} g^{b \nu} g^{c \alpha} g^{d \beta}  + B g^{ac} g^{\mu\alpha} g^{\nu b} g^{\beta d}  + Cg^{\nu\beta} g^{\mu\alpha} g^{bd} g^{ac}) R^\nabla_{\mu\nu\alpha\beta} R^\nabla_{abcd} .
\end{equation}

Since we require the Euler derivative for the Hilbert energy-momentum tensor $T^{\gamma\rho}_H = \frac{2}{\sqrt{-g}} \frac{\delta \mathcal{L}}{\delta  g_{\gamma\rho}} \Bigg|_{g = \eta}$, it is necessary to write $R^\nabla_{\mu\nu\alpha\beta}$ in terms of the metric and its derivatives. Therefore we require the second covariant derivatives of the tensor potential $h_{\nu\beta}$,

\begin{multline}
\nabla_\mu (\nabla_\alpha h_{\nu\beta}) = 
\partial_\mu (\partial_\alpha h_{\nu\beta} - \Gamma^\lambda_{\alpha \nu} h_{\lambda \beta} - \Gamma^\lambda_{\alpha \beta} h_{\nu \lambda}) 
- \Gamma^\lambda_{\mu \alpha} (\partial_\lambda h_{\nu\beta} - \Gamma^\rho_{\lambda \nu} h_{\rho \beta} - \Gamma^\rho_{\lambda \beta} h_{\nu \rho}) 
\\
- \Gamma^\lambda_{\mu \nu} (\partial_\alpha h_{\lambda\beta} - \Gamma^\rho_{\alpha \lambda} h_{\rho \beta} - \Gamma^\rho_{\alpha \beta} h_{\lambda \rho}) 
- \Gamma^\lambda_{\mu \beta} (\partial_\alpha h_{\nu\lambda} - \Gamma^\rho_{\alpha \nu} h_{\rho \lambda} - \Gamma^\rho_{\alpha \lambda} h_{\nu \rho}) ,
\end{multline}

where ${\Gamma}^\lambda_{\nu \beta}=\frac{1}{2} g^{\mu \lambda}( - \partial_\mu g_{\nu \beta}+  \partial_\beta g_{\mu \nu} +   \partial_\nu g_{\mu \beta})$ is the Christoffel symbol of the second kind. Since this term appears four times in $R^\nabla_{\mu\nu\alpha\beta}$, we are left with,

\begin{multline}
R^\nabla_{\mu\nu\alpha\beta} = \frac{1}{2} [
\partial_\mu (\partial_\alpha h_{\nu\beta} - \Gamma^\lambda_{\alpha \nu} h_{\lambda \beta} - \Gamma^\lambda_{\alpha \beta} h_{\nu \lambda}) 
- \Gamma^\lambda_{\mu \alpha} (\partial_\lambda h_{\nu\beta} - \Gamma^\rho_{\lambda \nu} h_{\rho \beta} - \Gamma^\rho_{\lambda \beta} h_{\nu \rho}) 
\\
- \Gamma^\lambda_{\mu \nu} (\partial_\alpha h_{\lambda\beta} - \Gamma^\rho_{\alpha \lambda} h_{\rho \beta} - \Gamma^\rho_{\alpha \beta} h_{\lambda \rho}) 
- \Gamma^\lambda_{\mu \beta} (\partial_\alpha h_{\nu\lambda} - \Gamma^\rho_{\alpha \nu} h_{\rho \lambda} - \Gamma^\rho_{\alpha \lambda} h_{\nu \rho})
 + 
\partial_\nu (\partial_\beta h_{\mu\alpha} - \Gamma^\lambda_{\beta \mu} h_{\lambda \alpha} - \Gamma^\lambda_{\beta \alpha} h_{\mu \lambda}) 
\\
- \Gamma^\lambda_{\nu \beta} (\partial_\lambda h_{\mu\alpha} - \Gamma^\rho_{\lambda \mu} h_{\rho \alpha} - \Gamma^\rho_{\lambda \alpha} h_{\mu \rho}) 
- \Gamma^\lambda_{\nu \mu} (\partial_\beta h_{\lambda\alpha} - \Gamma^\rho_{\beta \lambda} h_{\rho \alpha} - \Gamma^\rho_{\beta \alpha} h_{\lambda \rho}) 
- \Gamma^\lambda_{\nu \alpha} (\partial_\beta h_{\mu\lambda} - \Gamma^\rho_{\beta \mu} h_{\rho \lambda} - \Gamma^\rho_{\beta \lambda} h_{\mu \rho})
\\
 - \partial_\mu (\partial_\beta h_{\nu\alpha} - \Gamma^\lambda_{\beta \nu} h_{\lambda \alpha} - \Gamma^\lambda_{\beta \alpha} h_{\nu \lambda}) 
+ \Gamma^\lambda_{\mu \beta} (\partial_\lambda h_{\nu\alpha} - \Gamma^\rho_{\lambda \nu} h_{\rho \alpha} - \Gamma^\rho_{\lambda \alpha} h_{\nu \rho}) 
+ \Gamma^\lambda_{\mu \nu} (\partial_\beta h_{\lambda\alpha} - \Gamma^\rho_{\beta \lambda} h_{\rho \alpha} - \Gamma^\rho_{\beta \alpha} h_{\lambda \rho}) 
\\
+ \Gamma^\lambda_{\mu \alpha} (\partial_\beta h_{\nu\lambda} - \Gamma^\rho_{\beta \nu} h_{\rho \lambda} - \Gamma^\rho_{\beta \lambda} h_{\nu \rho})
 - \partial_\nu (\partial_\alpha h_{\mu\beta} - \Gamma^\lambda_{\alpha \mu} h_{\lambda \beta} - \Gamma^\lambda_{\alpha \beta} h_{\mu \lambda}) 
+ \Gamma^\lambda_{\nu \alpha} (\partial_\lambda h_{\mu\beta} - \Gamma^\rho_{\lambda \mu} h_{\rho \beta} - \Gamma^\rho_{\lambda \beta} h_{\mu \rho}) 
\\
+ \Gamma^\lambda_{\nu \mu} (\partial_\alpha h_{\lambda\beta} - \Gamma^\rho_{\alpha \lambda} h_{\rho \beta} - \Gamma^\rho_{\alpha \beta} h_{\lambda \rho}) 
+ \Gamma^\lambda_{\nu \beta} (\partial_\alpha h_{\mu\lambda} - \Gamma^\rho_{\alpha \mu} h_{\rho \lambda} - \Gamma^\rho_{\alpha \lambda} h_{\mu \rho}) ] . \label{rienab}
\end{multline}

Expanding this expression is a bit tedious. Many terms cancel, and combine. Familiar terms here, the Riemann tensor $\bar{R}^\rho_{\sigma\mu\nu} = \partial_\mu \Gamma^\rho_{\nu\sigma} - \partial_\nu \Gamma^\rho_{\mu\sigma} + \Gamma^\rho_{\mu\lambda} \Gamma^\lambda_{\nu\sigma} - \Gamma^\rho_{\nu\lambda} \Gamma^\lambda_{\mu\sigma}$ and linearized Christoffel symbol $\bar{\Gamma}_{\lambda \mu \alpha} = \frac{1}{2} (- \partial_\lambda h_{\mu\alpha} + \partial_\alpha h_{\lambda\mu} + \partial_\mu h_{\lambda \alpha} )$, allow for Equation (\ref{rienab}) to be expressed much more compactly as,

\begin{multline}
R^\nabla_{\mu\nu\alpha\beta} = 
{R}_{\mu\nu\alpha\beta} 
 - \frac{1}{2} \bar{R}^\lambda_{\alpha\mu\nu} h_{\lambda \beta}
+ \frac{1}{2} \bar{R}^\lambda_{\beta\mu\nu}h_{\lambda \alpha}
\\
- 2 \Gamma^\lambda_{\nu \alpha} \bar{\Gamma}_{\lambda \mu \beta}
- 2 \Gamma^\lambda_{\mu \beta} \bar{\Gamma}_{\lambda \nu \alpha}
+ 2 \Gamma^\lambda_{\mu \alpha} \bar{\Gamma}_{\lambda \nu \beta}
+ 2 \Gamma^\lambda_{\nu \beta} \bar{\Gamma}_{\lambda \mu \alpha}
+ \Gamma^\lambda_{\mu \beta}  \Gamma^\rho_{\alpha \nu} h_{\rho \lambda} 
-  \Gamma^\lambda_{\mu \alpha} \Gamma^\rho_{\beta \nu} h_{\rho \lambda} .
\end{multline}

We now make an important note that can save the reader many pages of calculations. The Hilbert energy-momentum tensor requires us to replace the metric tensor with the Minkowski metric after variation $g \to \eta$. Therefore any derivatives of the metric that remain after variation will be zero upon differentiation. Some terms in the Lagrangian, namely those of the form $\Gamma\Gamma$ (not $\bar{\Gamma}$ {\color{red} ,} because these are the linearized expressions) will all vanish upon $g \to \eta$. Due to this fact we will neglect such terms from $R^\nabla_{\mu\nu\alpha\beta}$, as they will not contribute to the final result. For clarity this will be labelled $R^{H \nabla }_{\mu\nu\alpha\beta}$ for the terms which contribute to the Hilbert energy-momentum tensor,

\begin{equation}
R^{H \nabla}_{\mu\nu\alpha\beta} = 
{R}_{\mu\nu\alpha\beta} 
 - \frac{1}{2} (\partial_\mu \Gamma^\lambda_{\nu\alpha} - \partial_\nu \Gamma^\lambda_{\mu\alpha})  h_{\lambda \beta}
+ \frac{1}{2} (\partial_\mu \Gamma^\lambda_{\nu\beta} - \partial_\nu \Gamma^\lambda_{\mu\beta}) h_{\lambda \alpha}\\
- 2 \Gamma^\lambda_{\nu \alpha} \bar{\Gamma}_{\lambda \mu \beta}
- 2 \Gamma^\lambda_{\mu \beta} \bar{\Gamma}_{\lambda \nu \alpha}
+ 2 \Gamma^\lambda_{\mu \alpha} \bar{\Gamma}_{\lambda \nu \beta}
+ 2 \Gamma^\lambda_{\nu \beta} \bar{\Gamma}_{\lambda \mu \alpha} .
\end{equation}

This expression can be further 'simplified' by noting that the $R^{H \nabla}_{\mu\nu\alpha\beta}$ is multiplied by $R^{H \nabla}_{abcd}$. The vast majority of terms in this expansion will have a $\Gamma \Gamma$ contribution. Therefore keeping only those which will be nonzero after variation for $R^{H \nabla}_{\mu\nu\alpha\beta} R^{H \nabla}_{abcd}$, we are left with the Lagrangian,

\begin{multline}
\mathcal{L} =  \sqrt{-g} (A g^{a \mu} g^{b \nu} g^{c \alpha} g^{d \beta}  + B g^{ac} g^{\mu\alpha} g^{\nu b} g^{\beta d}  + Cg^{\nu\beta} g^{\mu\alpha} g^{bd} g^{ac}) ({R}_{\mu\nu\alpha\beta} {R}_{abcd} 
\\
 - 2 \Gamma^\lambda_{\nu \alpha} {R}_{abcd}  \bar{\Gamma}_{\lambda \mu \beta}
- 2 \Gamma^\lambda_{\mu \beta} {R}_{abcd}  \bar{\Gamma}_{\lambda \nu \alpha}
+ 2 \Gamma^\lambda_{\mu \alpha} {R}_{abcd}  \bar{\Gamma}_{\lambda \nu \beta}
+ 2 \Gamma^\lambda_{\nu \beta} {R}_{abcd}  \bar{\Gamma}_{\lambda \mu \alpha}
\\
- 2 \Gamma^\gamma_{b c} {R}_{\mu\nu\alpha\beta}  \bar{\Gamma}_{\gamma a d}
- 2 \Gamma^\gamma_{a d} {R}_{\mu\nu\alpha\beta} \bar{\Gamma}_{\gamma b c}
+ 2 \Gamma^\gamma_{a c} {R}_{\mu\nu\alpha\beta} \bar{\Gamma}_{\gamma b d}
+ 2 \Gamma^\gamma_{b d} {R}_{\mu\nu\alpha\beta} \bar{\Gamma}_{\gamma a c}
\\
 - \frac{1}{2} \partial_\mu \Gamma^\lambda_{\nu\alpha} {R}_{abcd}  h_{\lambda \beta}
 +\frac{1}{2}  \partial_\nu \Gamma^\lambda_{\mu\alpha} {R}_{abcd}  h_{\lambda \beta}
+ \frac{1}{2} \partial_\mu \Gamma^\lambda_{\nu\beta} {R}_{abcd}  h_{\lambda \alpha}
- \frac{1}{2}  \partial_\nu \Gamma^\lambda_{\mu\beta} {R}_{abcd}  h_{\lambda \alpha}
\\
 - \frac{1}{2}  \partial_a \Gamma^\gamma_{b c} {R}_{\mu\nu\alpha\beta}  h_{\gamma d}
 + \frac{1}{2}   \partial_b \Gamma^\gamma_{ac} {R}_{\mu\nu\alpha\beta}  h_{\gamma d}
+ \frac{1}{2}  \partial_a \Gamma^\gamma_{b d}  {R}_{\mu\nu\alpha\beta}  h_{\gamma c}
- \frac{1}{2}   \partial_b \Gamma^\gamma_{ad} {R}_{\mu\nu\alpha\beta}  h_{\gamma c}) . \label{CurveLag}
\end{multline}

The Lagrangian terms are sorted as follows. The first line is terms which will be nonzero after differentiation $\frac{\partial \mathcal{L}}{\partial g_{\gamma\rho}} $ and $g \to \eta$, the second and third lines will be nonzero after $\partial_\omega \frac{\partial \mathcal{L}}{\partial (\partial_\omega g_{\gamma\rho})}$ and $g \to \eta$, the fourth and fifth lines will be nonzero after $\partial_\xi \partial_\omega \frac{\partial \mathcal{L}}{\partial (\partial_\xi \partial_\omega g_{\gamma\rho})}$ and $g \to \eta$. 

\subsection{Taking the Euler derivative of the Lagrangian}

Recall that we require the Euler derivative $\frac{\delta \mathcal{L}}{\delta  g_{\gamma\rho}} =  \frac{\partial \mathcal{L}}{\partial g_{\gamma\rho}} - \partial_\omega \frac{\partial \mathcal{L}}{\partial (\partial_\omega g_{\gamma\rho})} +  \partial_\xi \partial_\omega \frac{\partial \mathcal{L}}{\partial (\partial_\xi \partial_\omega g_{\gamma\rho})}$ in order to derive the Hilbert energy-momentum tensor. Performing the differentiation of the  relevant parts, first with respect to the metric, the nonzero terms after $g \to \eta$ are,

\begin{multline}
 \frac{\partial \mathcal{L}}{\partial g_{\gamma\rho}}  = 
 \frac{\partial  \sqrt{-g}}{\partial g_{\gamma\rho}} 
[{A}g^{\mu a} g^{\nu b} g^{\alpha c} g^{\beta d}
+  {B}    g^{\nu b} g^{\beta d} g^{\mu\alpha} g^{ac}
+ {C}  g^{\mu\alpha} g^{\nu\beta} g^{ac} g^{bd}]  (  {R}_{\mu\nu\alpha\beta} {R}_{abcd}  )
\\
+    \sqrt{-g}  [{A} \frac{\partial g^{\mu a}}{\partial g_{\gamma\rho}}  g^{\nu b} g^{\alpha c} g^{\beta d} 
+ {A} g^{\mu a} \frac{\partial g^{\nu b}}{\partial g_{\gamma\rho}} g^{\alpha c} g^{\beta d} 
+ {A} g^{\mu a} g^{\nu b} \frac{\partial g^{\alpha c}}{\partial g_{\gamma\rho}} g^{\beta d} 
+ {A} g^{\mu a} g^{\nu b} g^{\alpha c} \frac{\partial g^{\beta d}}{\partial g_{\gamma\rho}}
\\
+  {B}  \frac{\partial g^{\nu b}}{\partial g_{\gamma\rho}}  g^{\beta d} g^{\mu\alpha} g^{ac}
+ {B}   g^{\nu b}   \frac{\partial g^{\beta d}}{\partial g_{\gamma\rho}}  g^{\mu\alpha} g^{ac}
+ {B}   g^{\nu b} g^{\beta d}   \frac{\partial g^{\mu\alpha}}{\partial g_{\gamma\rho}}  g^{ac}
+ {B}   g^{\nu b} g^{\beta d} g^{\mu\alpha}   \frac{\partial g^{ac}}{\partial g_{\gamma\rho}} 
\\
+ {C}  \frac{\partial g^{\mu\alpha}}{\partial g_{\gamma\rho}}  g^{\nu\beta} g^{ac} g^{bd}
+ {C} g^{\mu\alpha}   \frac{\partial g^{\nu\beta}}{\partial g_{\gamma\rho}}  g^{ac} g^{bd}
+ {C} g^{\mu\alpha} g^{\nu\beta}   \frac{\partial g^{ac}}{\partial g_{\gamma\rho}}  g^{bd}
+ {C} g^{\mu\alpha} g^{\nu\beta} g^{ac}   \frac{\partial g^{bd}}{\partial g_{\gamma\rho}} ]  (  {R}_{\mu\nu\alpha\beta} {R}_{abcd}  ) .
\end{multline}

Inserting $ \frac{\partial \sqrt{-g}}{\partial g_{\gamma\rho}}$ and $ \frac{\partial g^{\lambda\nu}}{ \partial g_{\beta\gamma}}$, and expanding all brackets, we are left with the following expression,

\begin{multline}
 \frac{\partial \mathcal{L}}{\partial g_{\gamma\rho}}  = 
  -  \frac{1}{2}  \sqrt{-g}
(- {A} g^{\gamma\rho} g^{\mu a} g^{\nu b} g^{\alpha c} g^{\beta d}
+ {A} g^{\gamma\mu} g^{\rho a} g^{\nu b} g^{\alpha c} g^{\beta d}  
+   {A} g^{\rho\mu} g^{\gamma a} g^{\nu b} g^{\alpha c} g^{\beta d} 
\\
+  {A} g^{\gamma b} g^{\rho \nu}  g^{\mu a}  g^{\alpha c} g^{\beta d} 
+   {A} g^{\rho b} g^{\gamma \nu} g^{\mu a}  g^{\alpha c} g^{\beta d} 
+  {A} g^{\gamma c} g^{\rho \alpha} g^{\mu a} g^{\nu b}  g^{\beta d}  
+   {A} g^{\rho c} g^{\gamma \alpha} g^{\mu a} g^{\nu b}  g^{\beta d} 
\\
+  {A} g^{\gamma d} g^{\rho \beta}  g^{\mu a} g^{\nu b} g^{\alpha c} 
+   {A} g^{\rho d} g^{\gamma \beta} g^{\mu a} g^{\nu b} g^{\alpha c}
-  {B}   g^{\gamma\rho}   g^{\nu b} g^{\beta d} g^{\mu\alpha} g^{ac}
+   {B} g^{\gamma b} g^{\rho \nu} g^{\beta d} g^{\mu\alpha} g^{ac} 
\\
+    {B} g^{\rho b} g^{\gamma \nu} g^{\beta d} g^{\mu\alpha} g^{ac}
+   {B} g^{\gamma d} g^{\rho \beta} g^{\nu b}    g^{\mu\alpha} g^{ac} 
+    {B} g^{\rho d} g^{\gamma \beta} g^{\nu b}    g^{\mu\alpha} g^{ac}
+   {B} g^{\gamma\mu} g^{\rho \alpha} g^{\nu b} g^{\beta d}  g^{ac} 
\\
+    {B} g^{\rho\mu} g^{\gamma \alpha} g^{\nu b} g^{\beta d}  g^{ac}
+   {B} g^{\gamma c} g^{\rho a} g^{\nu b} g^{\beta d} g^{\mu\alpha}  
+    {B} g^{\rho c} g^{\gamma a} g^{\nu b} g^{\beta d} g^{\mu\alpha} 
- {C}   g^{\gamma\rho} g^{\mu\alpha} g^{\nu\beta} g^{ac} g^{bd}
\\
+  {C} g^{\gamma\mu} g^{\rho \alpha}  g^{\nu\beta} g^{ac} g^{bd} 
+   {C} g^{\rho\mu} g^{\gamma \alpha}  g^{\nu\beta} g^{ac} g^{bd} 
+  {C} g^{\gamma \beta} g^{\rho \nu} g^{\mu\alpha} g^{ac} g^{bd} 
+   {C} g^{\rho \beta} g^{\gamma \nu} g^{\mu\alpha} g^{ac} g^{bd}
\\
+ {C} g^{\gamma c} g^{\rho a} g^{\mu\alpha} g^{\nu\beta}    g^{bd} 
+   {C} g^{\rho c} g^{\gamma a} g^{\mu\alpha} g^{\nu\beta}    g^{bd}
+  {C} g^{\gamma d} g^{\rho b} g^{\mu\alpha} g^{\nu\beta} g^{ac}   
+  {C}  g^{\rho d} g^{\gamma b} g^{\mu\alpha} g^{\nu\beta} g^{ac}  )
(  {R}_{\mu\nu\alpha\beta} {R}_{abcd}  ) .
\end{multline}

To write this expression more compactly we express the terms proportional to $A$ as $\bar{g}_A^{\gamma\rho\mu a \nu b \alpha c \beta d} = 
-  g^{\gamma\rho} g^{\mu a} g^{\nu b} g^{\alpha c} g^{\beta d}
+ g^{\gamma\mu} g^{\rho a} g^{\nu b} g^{\alpha c} g^{\beta d} 
+ g^{\rho\mu} g^{\gamma a} g^{\nu b} g^{\alpha c} g^{\beta d} 
+ g^{\gamma b} g^{\rho \nu}  g^{\mu a}  g^{\alpha c} g^{\beta d}
+   g^{\rho b} g^{\gamma \nu} g^{\mu a}  g^{\alpha c} g^{\beta d} 
+  g^{\gamma c} g^{\rho \alpha} g^{\mu a} g^{\nu b}  g^{\beta d}  
+  g^{\rho c} g^{\gamma \alpha} g^{\mu a} g^{\nu b}  g^{\beta d} 
+  g^{\gamma d} g^{\rho \beta}  g^{\mu a} g^{\nu b} g^{\alpha c} 
+   g^{\rho d} g^{\gamma \beta} g^{\mu a} g^{\nu b} g^{\alpha c}$,

we express the terms proportional to $B$ as $\bar{g}_B^{\gamma\rho \nu b \beta d \mu\alpha ac} = 
-   g^{\gamma\rho}   g^{\nu b} g^{\beta d} g^{\mu\alpha} g^{ac}
+ g^{\gamma b} g^{\rho \nu} g^{\beta d} g^{\mu\alpha} g^{ac} 
+     g^{\rho b} g^{\gamma \nu} g^{\beta d} g^{\mu\alpha} g^{ac}
+   g^{\gamma d} g^{\rho \beta} g^{\nu b}    g^{\mu\alpha} g^{ac} 
+   g^{\rho d} g^{\gamma \beta} g^{\nu b}    g^{\mu\alpha} g^{ac}
+  g^{\gamma\mu} g^{\rho \alpha} g^{\nu b} g^{\beta d}  g^{ac} 
+    g^{\rho\mu} g^{\gamma \alpha} g^{\nu b} g^{\beta d}  g^{ac}
+  g^{\gamma c} g^{\rho a} g^{\nu b} g^{\beta d} g^{\mu\alpha}  
+    g^{\rho c} g^{\gamma a} g^{\nu b} g^{\beta d} g^{\mu\alpha} $

and we express the terms proportional to $C$ as $\bar{g}_C^{\gamma\rho \mu\alpha \nu\beta ac bd} = 
-  g^{\gamma\rho} g^{\mu\alpha} g^{\nu\beta} g^{ac} g^{bd}
+  g^{\gamma\mu} g^{\rho \alpha}  g^{\nu\beta} g^{ac} g^{bd} 
+   g^{\rho\mu} g^{\gamma \alpha}  g^{\nu\beta} g^{ac} g^{bd} 
+   g^{\gamma \beta} g^{\rho \nu} g^{\mu\alpha} g^{ac} g^{bd} 
+   g^{\rho \beta} g^{\gamma \nu} g^{\mu\alpha} g^{ac} g^{bd}
+  g^{\gamma c} g^{\rho a} g^{\mu\alpha} g^{\nu\beta}    g^{bd} 
+    g^{\rho c} g^{\gamma a} g^{\mu\alpha} g^{\nu\beta}    g^{bd}
+  g^{\gamma d} g^{\rho b} g^{\mu\alpha} g^{\nu\beta} g^{ac}   
+   g^{\rho d} g^{\gamma b} g^{\mu\alpha} g^{\nu\beta} g^{ac} $. 

Therefore the derivative of the Lagrangian with respect to the metric is expressed compactly as,

\begin{equation}
 \frac{\partial \mathcal{L}}{\partial g_{\gamma\rho}}  = 
  -  \frac{1}{2}  \sqrt{-g}
({A} \bar{g}_A^{\gamma\rho\mu a \nu b \alpha c \beta d}
+  {B}   \bar{g}_B^{\gamma\rho \nu b \beta d \mu\alpha ac}
+ {C}   \bar{g}_C^{\gamma\rho \mu\alpha \nu\beta ac bd} )
 {R}_{\mu\nu\alpha\beta} {R}_{abcd}   . \label{euldivt1} 
\end{equation}

Next we will differentiate the Lagrangian in Equation (\ref{CurveLag}) with respect to derivatives of the metric. Only the second and third lines of the Lagrangian in Equation (\ref{CurveLag}) will be nonzero after $ \frac{\partial \mathcal{L}}{\partial (\partial_\omega g_{\gamma\rho})}$, as only terms with a non-linearized Christoffel symbol ${\Gamma}^\lambda_{\nu \alpha}=\frac{1}{2} g^{m \lambda}( - \partial_m g_{\nu \alpha}+  \partial_\alpha g_{m \nu} +   \partial_\nu g_{m \alpha})$ have linear in $\partial g$ contributions. These terms will be differentiating as $\frac{\partial  \partial_m g_{\nu \alpha}}{\partial (\partial_\omega g_{\gamma\rho})} = \delta^\omega_m \Delta^{\gamma \rho}_{\nu \alpha}$ where $\Delta^{\gamma \rho}_{\nu \alpha} = \frac{1}{2} (\delta^{\gamma }_{\nu } \delta^{ \rho}_{ \alpha} + \delta^{\gamma }_{ \alpha} \delta^{ \rho}_{\nu })$. Differentiating the Chrisoffel symbol therefore yields,

\begin{equation}
 \frac{\partial {\Gamma}^\lambda_{\nu \alpha}}{\partial (\partial_\omega g_{\gamma\rho})} = \frac{1}{2} g^{m \lambda}
( -  \delta^\omega_m \Delta^{\gamma \rho}_{\nu \alpha}  
+  \delta^\omega_\alpha \Delta^{\gamma \rho}_{m \nu}
 + \delta^\omega_\nu \Delta^{\gamma \rho}_{m \alpha} ) = \frac{1}{2} g^{m \lambda} \bar{\Delta}^{\omega\gamma\rho}_{m \nu\alpha} ,
\end{equation}

where above to abbreviate we call the combination of the Kronecker deltas in brackets $\bar{\Delta}^{\omega\gamma\rho}_{m \nu\alpha} =  -  \delta^\omega_m \Delta^{\gamma \rho}_{\nu \alpha}  +  \delta^\omega_\alpha \Delta^{\gamma \rho}_{m \nu} + \delta^\omega_\nu \Delta^{\gamma \rho}_{m \alpha}$. Using this compact notiation the derivative of the Lagrangian with respect to derivatives of the metric is,

\begin{multline}
 \frac{\partial \mathcal{L}}{\partial (\partial_\omega g_{\gamma\rho})}=  \sqrt{-g} (A g^{a \mu} g^{b \nu} g^{c \alpha} g^{d \beta}  + B g^{ac} g^{\mu\alpha} g^{\nu b} g^{\beta d}  + Cg^{\nu\beta} g^{\mu\alpha} g^{bd} g^{ac}) (
\\
 - g^{m \lambda} \bar{\Delta}^{\omega\gamma\rho}_{m \nu\alpha} {R}_{abcd}  \bar{\Gamma}_{\lambda \mu \beta}
-  g^{m \lambda} \bar{\Delta}^{\omega\gamma\rho}_{m \mu\beta} {R}_{abcd}  \bar{\Gamma}_{\lambda \nu \alpha}
+ g^{m \lambda} \bar{\Delta}^{\omega\gamma\rho}_{m \mu\alpha} {R}_{abcd}  \bar{\Gamma}_{\lambda \nu \beta}
+ g^{m \lambda} \bar{\Delta}^{\omega\gamma\rho}_{m \nu\beta} {R}_{abcd}  \bar{\Gamma}_{\lambda \mu \alpha}
\\
- g^{m \lambda} \bar{\Delta}^{\omega\gamma\rho}_{m b c} {R}_{\mu\nu\alpha\beta}  \bar{\Gamma}_{\lambda a d}
- g^{m \lambda} \bar{\Delta}^{\omega\gamma\rho}_{m a d} {R}_{\mu\nu\alpha\beta} \bar{\Gamma}_{\lambda b c}
+ g^{m \lambda} \bar{\Delta}^{\omega\gamma\rho}_{m a c} {R}_{\mu\nu\alpha\beta} \bar{\Gamma}_{\lambda b d}
+ g^{m \lambda} \bar{\Delta}^{\omega\gamma\rho}_{m b d}{R}_{\mu\nu\alpha\beta} \bar{\Gamma}_{\lambda a c}
) . \label{dgderiv}
\end{multline}

Since we require $\partial_\omega \frac{\partial \mathcal{L}}{\partial (\partial_\omega g_{\gamma\rho})}$ we must differentiate the above expression with by $\partial_\omega$. This process is in general quite messy, but since any $\partial g$ will be zero upon $g \to \eta$, only the linearized Riemann tensor and linearized Christoffel symbol will, differentiated, give rise to nonzero contributions,

\begin{multline}
\partial_\omega  \frac{\partial \mathcal{L}}{\partial (\partial_\omega g_{\gamma\rho})}=  2 \sqrt{-g} (A g^{a \mu} g^{b \nu} g^{c \alpha} g^{d \beta}  + B g^{ac} g^{\mu\alpha} g^{\nu b} g^{\beta d}  + Cg^{\nu\beta} g^{\mu\alpha} g^{bd} g^{ac}) (
\\
 - g^{m \lambda} \bar{\Delta}^{\omega\gamma\rho}_{m \nu\alpha} \partial_\omega [{R}_{abcd}  \bar{\Gamma}_{\lambda \mu \beta}]
-  g^{m \lambda} \bar{\Delta}^{\omega\gamma\rho}_{m \mu\beta} \partial_\omega [{R}_{abcd}  \bar{\Gamma}_{\lambda \nu \alpha}]
+ g^{m \lambda} \bar{\Delta}^{\omega\gamma\rho}_{m \mu\alpha} \partial_\omega [{R}_{abcd}  \bar{\Gamma}_{\lambda \nu \beta}]
+ g^{m \lambda} \bar{\Delta}^{\omega\gamma\rho}_{m \nu\beta} \partial_\omega [{R}_{abcd}  \bar{\Gamma}_{\lambda \mu \alpha}]
) , \label{euldivt2}
\end{multline}

where the final two lines in Equation (\ref{dgderiv}) were combined by interchange $abcd \leftrightarrow \mu\nu\alpha\beta$. Finally for the terms proportional to $\partial \partial g$ in the fourth and fifth lines of the Lagrangian in Equation (\ref{CurveLag}), we require the differentiated Christoffel symbol $\partial_a {\Gamma}^\lambda_{b c}=\frac{1}{2} \partial_a g^{m \lambda}( - \partial_m g_{b c}+  \partial_c g_{m b} +   \partial_b g_{m c})+ \frac{1}{2} g^{m \lambda} \partial_a ( - \partial_m g_{b c}+  \partial_c g_{m b} +   \partial_b g_{m c})$. The first term will be zero upon $g \to \eta$ so we can neglect it, leaving $\partial_a {\Gamma}^\lambda_{b c}= \frac{1}{2} g^{m \lambda} ( -  \partial_a \partial_m g_{b c}+   \partial_a \partial_c g_{m b} +    \partial_a \partial_b g_{m c})$. Differentiating each term will yields $\frac{\partial \partial_a \partial_m g_{b c}}{\partial (\partial_\xi \partial_\omega g_{\gamma\rho})} = \Delta^{\xi \omega}_{am} \Delta^{\gamma \rho}_{bc}$. Therefore differentiating of the derivative of the Christoffel symbol gives,

\begin{equation}
\frac{\partial     \partial_a {\Gamma}^\lambda_{b c} }{\partial (\partial_\xi \partial_\omega g_{\gamma\rho})}
=
 \frac{1}{2} g^{m \lambda} (- \Delta^{\xi \omega}_{am} \Delta^{\gamma \rho}_{bc}
+  \Delta^{\xi \omega}_{ac} \Delta^{\gamma \rho}_{mb}
 +  \Delta^{\xi \omega}_{ab} \Delta^{\gamma \rho}_{mc}) =  \frac{1}{2} g^{m \lambda} \hat{\Delta}^{\xi\omega\gamma\rho}_{ambc} .
\end{equation}

The above expression in brackets was abbreviated with $\hat{\Delta}^{\xi\omega\gamma\rho}_{ambc} = - \Delta^{\xi \omega}_{am} \Delta^{\gamma \rho}_{bc} +  \Delta^{\xi \omega}_{ac} \Delta^{\gamma \rho}_{mb} +  \Delta^{\xi \omega}_{ab} \Delta^{\gamma \rho}_{mc}$. Using this compact notation the derivative of the Lagrangian with respect to two derivatives of the metric is,

\begin{multline}
\frac{\partial    \mathcal{L}}{\partial (\partial_\xi \partial_\omega g_{\gamma\rho})}  = \frac{1}{4}  \sqrt{-g} (A g^{a \mu} g^{b \nu} g^{c \alpha} g^{d \beta}  + B g^{ac} g^{\mu\alpha} g^{\nu b} g^{\beta d}  + Cg^{\nu\beta} g^{\mu\alpha} g^{bd} g^{ac}) (
\\
 - g^{m \lambda} \hat{\Delta}^{\xi\omega\gamma\rho}_{\mu m \nu \alpha} {R}_{abcd}  h_{\lambda \beta}
 + g^{m \lambda} \hat{\Delta}^{\xi\omega\gamma\rho}_{\nu m \mu \alpha} {R}_{abcd}  h_{\lambda \beta}
+ g^{m \lambda} \hat{\Delta}^{\xi\omega\gamma\rho}_{\mu m \nu \beta} {R}_{abcd}  h_{\lambda \alpha}
-  g^{m \lambda} \hat{\Delta}^{\xi\omega\gamma\rho}_{\nu m \mu \beta} {R}_{abcd}  h_{\lambda \alpha}
\\
 -   g^{m \lambda} \hat{\Delta}^{\xi\omega\gamma\rho}_{ambc} {R}_{\mu\nu\alpha\beta}  h_{\lambda d}
 +  g^{m \lambda} \hat{\Delta}^{\xi\omega\gamma\rho}_{bmac} {R}_{\mu\nu\alpha\beta}  h_{\lambda d}
+  g^{m \lambda} \hat{\Delta}^{\xi\omega\gamma\rho}_{ambd}  {R}_{\mu\nu\alpha\beta}  h_{\lambda c}
-    g^{m \lambda} \hat{\Delta}^{\xi\omega\gamma\rho}_{bmad}  {R}_{\mu\nu\alpha\beta}  h_{\lambda c}) .
\end{multline}

Since we require $\partial_\xi \partial_\omega \frac{\partial    \mathcal{L}}{\partial (\partial_\xi \partial_\omega g_{\gamma\rho})}$ we must differentiate the above expression by $\partial_\xi \partial_\omega$. Again this process is in general quite messy, but since any $\partial g$ will be zero upon $g \to \eta$, only the linearized Riemann tensor and the potential $h_{\lambda \alpha}$, differentiated, give rise to nonzero contributions,

\begin{multline}
\partial_\xi \partial_\omega \frac{\partial    \mathcal{L}}{\partial (\partial_\xi \partial_\omega g_{\gamma\rho})}  = \frac{1}{2}  \sqrt{-g} (A g^{a \mu} g^{b \nu} g^{c \alpha} g^{d \beta}  + B g^{ac} g^{\mu\alpha} g^{\nu b} g^{\beta d}  + Cg^{\nu\beta} g^{\mu\alpha} g^{bd} g^{ac}) (
\\
 -   g^{m \lambda} \hat{\Delta}^{\xi\omega\gamma\rho}_{ambc} \partial_\xi \partial_\omega [{R}_{\mu\nu\alpha\beta}  h_{\lambda d}]
 +  g^{m \lambda} \hat{\Delta}^{\xi\omega\gamma\rho}_{bmac} \partial_\xi \partial_\omega [{R}_{\mu\nu\alpha\beta}  h_{\lambda d}]
\\
+  g^{m \lambda} \hat{\Delta}^{\xi\omega\gamma\rho}_{ambd}  \partial_\xi \partial_\omega [{R}_{\mu\nu\alpha\beta}  h_{\lambda c}]
-    g^{m \lambda} \hat{\Delta}^{\xi\omega\gamma\rho}_{bmad}  \partial_\xi \partial_\omega [{R}_{\mu\nu\alpha\beta}  h_{\lambda c}]) , \label{euldivt3}
\end{multline}

where the bottom two lines in $\frac{\partial    \mathcal{L}}{\partial (\partial_\xi \partial_\omega g_{\gamma\rho})}$ were combined by interchange $abcd \leftrightarrow \mu\nu\alpha\beta$. Therefore for the total Euler derivative we have, combining equations (\ref{euldivt1}), (\ref{euldivt2}) and (\ref{euldivt3}),

\begin{multline}
\frac{\delta \mathcal{L}}{\delta  g_{\gamma\rho}} =  - \frac{1}{2} \sqrt{-g}  \Bigg(  
({A} \bar{g}_A^{\gamma\rho\mu a \nu b \alpha c \beta d}
+  {B}   \bar{g}_B^{\gamma\rho \nu b \beta d \mu\alpha ac}
+ {C}   \bar{g}_C^{\gamma\rho \mu\alpha \nu\beta ac bd} )
  {R}_{\mu\nu\alpha\beta} {R}_{abcd}  
\\
+ 4 (A g^{a \mu} g^{b \nu} g^{c \alpha} g^{d \beta}  + B g^{ac} g^{\mu\alpha} g^{\nu b} g^{\beta d}  + Cg^{\nu\beta} g^{\mu\alpha} g^{bd} g^{ac}) (
- g^{m \lambda} \bar{\Delta}^{\omega\gamma\rho}_{m b c} \partial_\omega [{R}_{\mu\nu\alpha\beta}  \bar{\Gamma}_{\lambda a d}]
\\
- g^{m \lambda} \bar{\Delta}^{\omega\gamma\rho}_{m a d} \partial_\omega [{R}_{\mu\nu\alpha\beta} \bar{\Gamma}_{\lambda b c}]
+ g^{m \lambda} \bar{\Delta}^{\omega\gamma\rho}_{m a c} \partial_\omega [{R}_{\mu\nu\alpha\beta} \bar{\Gamma}_{\lambda b d}]
+ g^{m \lambda} \bar{\Delta}^{\omega\gamma\rho}_{m b d} \partial_\omega [{R}_{\mu\nu\alpha\beta} \bar{\Gamma}_{\lambda a c}]
)
\\
-  (A g^{a \mu} g^{b \nu} g^{c \alpha} g^{d \beta}  + B g^{ac} g^{\mu\alpha} g^{\nu b} g^{\beta d}  + Cg^{\nu\beta} g^{\mu\alpha} g^{bd} g^{ac}) (
 -   g^{m \lambda} \hat{\Delta}^{\xi\omega\gamma\rho}_{ambc} \partial_\xi \partial_\omega [{R}_{\mu\nu\alpha\beta}  h_{\lambda d}]
\\
 +  g^{m \lambda} \hat{\Delta}^{\xi\omega\gamma\rho}_{bmac} \partial_\xi \partial_\omega [{R}_{\mu\nu\alpha\beta}  h_{\lambda d}]
+  g^{m \lambda} \hat{\Delta}^{\xi\omega\gamma\rho}_{ambd}  \partial_\xi \partial_\omega [{R}_{\mu\nu\alpha\beta}  h_{\lambda c}]
-    g^{m \lambda} \hat{\Delta}^{\xi\omega\gamma\rho}_{bmad}  \partial_\xi \partial_\omega [{R}_{\mu\nu\alpha\beta}  h_{\lambda c}]) \Bigg)  .
\end{multline}

\subsection{The Hilbert energy-momentum tensor}

We can now turn our attention to the Hilbert energy-momentum tensor $T^{\gamma\rho}_H$ in Equation (\ref{genhilbert}). Since we have calculated the Euler derivative in the previous section, evaluating this expression for $g \to \eta$ yields,

\begin{multline}
T^{\gamma\rho}_H =  
- ({A} \bar{\eta}_A^{\gamma\rho\mu a \nu b \alpha c \beta d}
+  {B}   \bar{\eta}_B^{\gamma\rho \nu b \beta d \mu\alpha ac}
+ {C}   \bar{\eta}_C^{\gamma\rho \mu\alpha \nu\beta ac bd} )
  {R}_{\mu\nu\alpha\beta} {R}_{abcd}  
\\
- 4 (A \eta^{a \mu} \eta^{b \nu} \eta^{c \alpha} \eta^{d \beta}  + B \eta^{ac} \eta^{\mu\alpha} \eta^{\nu b} \eta^{\beta d}  + C \eta^{\nu\beta} \eta^{\mu\alpha} \eta^{bd} \eta^{ac}) (
- \eta^{m \lambda} \bar{\Delta}^{\omega\gamma\rho}_{m b c} \partial_\omega [{R}_{\mu\nu\alpha\beta}  \bar{\Gamma}_{\lambda a d}]
\\
- \eta^{m \lambda} \bar{\Delta}^{\omega\gamma\rho}_{m a d} \partial_\omega [{R}_{\mu\nu\alpha\beta} \bar{\Gamma}_{\lambda b c}]
+ \eta^{m \lambda} \bar{\Delta}^{\omega\gamma\rho}_{m a c} \partial_\omega [{R}_{\mu\nu\alpha\beta} \bar{\Gamma}_{\lambda b d}]
+ \eta^{m \lambda} \bar{\Delta}^{\omega\gamma\rho}_{m b d} \partial_\omega [{R}_{\mu\nu\alpha\beta} \bar{\Gamma}_{\lambda a c}]
)
\\
+  (A \eta^{a \mu} \eta^{b \nu} \eta^{c \alpha} \eta^{d \beta}  + B \eta^{ac} \eta^{\mu\alpha} \eta^{\nu b} \eta^{\beta d}  + C \eta^{\nu\beta} \eta^{\mu\alpha} \eta^{bd} \eta^{ac}) (
 -   \eta^{m \lambda} \hat{\Delta}^{\xi\omega\gamma\rho}_{ambc} \partial_\xi \partial_\omega [{R}_{\mu\nu\alpha\beta}  h_{\lambda d}]
\\
 +  \eta^{m \lambda} \hat{\Delta}^{\xi\omega\gamma\rho}_{bmac} \partial_\xi \partial_\omega [{R}_{\mu\nu\alpha\beta}  h_{\lambda d}]
+  \eta^{m \lambda} \hat{\Delta}^{\xi\omega\gamma\rho}_{ambd}  \partial_\xi \partial_\omega [{R}_{\mu\nu\alpha\beta}  h_{\lambda c}]
-    \eta^{m \lambda} \hat{\Delta}^{\xi\omega\gamma\rho}_{bmad}  \partial_\xi \partial_\omega [{R}_{\mu\nu\alpha\beta}  h_{\lambda c}])  ,
\end{multline}

where the $\bar{g}_A^{\gamma\rho\mu a \nu b \alpha c \beta d} \to \bar{\eta}_A^{\gamma\rho\mu a \nu b \alpha c \beta d}$ is the same form with the metric tensor replaced by the Minkowski metric. If we contract all of the Minkowski tensors and the $\bar{\eta}_A^{\gamma\rho\mu a \nu b \alpha c \beta d}$, then we obtain,

\begin{multline}
T^{\gamma\rho}_H =  
-  {A} (- \eta^{\gamma \rho} R^{\a \b \c \d} R_{\a \b \c \d} + 8 R^{\gamma \b \c \d} R_{\ \b \c \d}^{\rho} ) 
-   {B}  (- \eta^{\gamma \rho} R^{ \b \d} R_{\b \d}+ 4 R^{ \gamma \d} R_{\d}^{ \rho} 
+ 4 R^{\gamma \b \rho \d} R_{\b \d} )
- {C} (- \eta^{\gamma \rho} R^2 + 8 R^{\gamma  \rho} R )
\\
- 16 A   \bar{\Delta}^{\omega\gamma\rho}_{m a c} \partial_\omega [{R}^{\a\b\c\d} \bar{\Gamma}^m_{\ b d}]
- 8 C  (
-  \bar{\Delta}^{\omega\gamma\rho}_{m a d} \partial_\omega [{R}  \bar{\Gamma}^{m \d\a}]
+  \eta^{ad} \bar{\Delta}^{\omega\gamma\rho}_{m a d} \partial_\omega [{R} \bar{\Gamma}^{m \b}_{\ \ \ b}])
\\
- 4 B (
- \bar{\Delta}^{\omega\gamma\rho}_{m b a} \partial_\omega [{R}^{\b\d}  \bar{\Gamma}^{m \a}_{\ \ \ d}]
-  \bar{\Delta}^{\omega\gamma\rho}_{m a d} \partial_\omega [{R}^{\b\d} \bar{\Gamma}^{m \ \a}_{\ \ b \ }]
+  \eta^{\a\c} \bar{\Delta}^{\omega\gamma\rho}_{m a c} \partial_\omega [{R}^{\b\d} \bar{\Gamma}^m_{\ b d}]
+  \bar{\Delta}^{\omega\gamma\rho}_{m b d} \partial_\omega [{R}^{\b\d} \bar{\Gamma}^{m \c}_{\ \ \ c}] )
\\
+ 4  A  \hat{\Delta}^{\xi\omega\gamma\rho}_{ambd}  \partial_\xi \partial_\omega [{R}^{\a\b\c\d}  h^m_{\ c}]
+  2 C  (
 - \eta^{ac} \hat{\Delta}^{\xi\omega\gamma\rho}_{ambc} \partial_\xi \partial_\omega [{R}  h^{m \b}]
 + \eta^{ac}  \hat{\Delta}^{\xi\omega\gamma\rho}_{bmac} \partial_\xi \partial_\omega [{R}  h^{m \b}])
\\
+  B (
 -   \eta^{ac}  \hat{\Delta}^{\xi\omega\gamma\rho}_{ambc} \partial_\xi \partial_\omega [{R}^{\b\d}   h^m_{\ d}]
 +   \eta^{ac}  \hat{\Delta}^{\xi\omega\gamma\rho}_{bmac} \partial_\xi \partial_\omega [{R}^{\b\d}  h^m_{\ d}]
+ \hat{\Delta}^{\xi\omega\gamma\rho}_{ambd}  \partial_\xi \partial_\omega [{R}^{\b\d}  h^{m \a}]
-   \hat{\Delta}^{\xi\omega\gamma\rho}_{bmad}  \partial_\xi \partial_\omega [{R}^{\b\d}  h^{m \a}]) .
\end{multline}

Next we will contract all of the $\bar{\Delta}^{\omega\gamma\rho}_{m b a}$ and $\hat{\Delta}^{\xi\omega\gamma\rho}_{ambd}$,

\begin{multline}
T^{\gamma\rho}_H =  
-  {A} (- \eta^{\gamma \rho} R^{\a \b \c \d} R_{\a \b \c \d} + 8 R^{\gamma \b \c \d} R_{\ \b \c \d}^{\rho} ) 
-   {B}  (- \eta^{\gamma \rho} R^{ \b \d} R_{\b \d}+ 4 R^{ \gamma \d} R_{\d}^{ \rho} 
+ 4 R^{\gamma \b \rho \d} R_{\b \d} )
\\
- {C} (- \eta^{\gamma \rho} R^2 + 8 R^{\gamma  \rho} R )
+ 16 A   \partial_\omega [R^{\gamma\b\rho\d} \bar{\Gamma}^\omega_{\ \b \d}
 - R^{\rho\b\omega\d} \bar{\Gamma}^\gamma_{\ \b \d}
 - R^{\gamma\b\omega\d} \bar{\Gamma}^\rho_{\ \b \d}]
\\
+ 4 B   \partial_\omega [- R^{\rho\d}  \bar{\Gamma}^{\omega \gamma}_{\ \ \ \d}
+ \eta^{\gamma\rho}  R^{\b\d} \bar{\Gamma}^\omega_{\ \b \d}
+ R^{\gamma\rho} \bar{\Gamma}^{\omega \b}_{\ \ \ \b}
- R^{\gamma\d}  \bar{\Gamma}^{\omega \rho}_{\ \ \ \d}
+ R^{\omega\d}  \bar{\Gamma}^{\gamma \rho}_{\ \ \ \d} 
 - \eta^{\omega\rho} R^{\b\d} \bar{\Gamma}^\gamma_{\ \b \d}] 
\\
 + 4 B  \partial_\omega [- R^{\rho\omega} \bar{\Gamma}^{\gamma \b}_{\ \ \ \b}
+ R^{\omega\d}  \bar{\Gamma}^{\rho \gamma}_{\ \ \ \d}
-  \eta^{\omega\gamma} R^{\b\d} \bar{\Gamma}^\rho_{\ \b \d}
-  R^{\gamma\omega} \bar{\Gamma}^{\rho \b}_{\ \ \ \b} 
+ R^{\rho\d}  \bar{\Gamma}^{\gamma \omega}_{\ \ \ \d}
+ R^{\gamma\d}  \bar{\Gamma}^{\rho \omega}_{\ \ \ \d}]  
\\
+ 8 C \partial_\omega [- R  \bar{\Gamma}^{\omega \rho\gamma}
+ \eta^{\gamma \rho} R \bar{\Gamma}^{\omega \b}_{\ \ \ \b} 
+ R  \bar{\Gamma}^{\gamma \omega\rho}
- \eta^{\omega\rho} R \bar{\Gamma}^{\gamma \b}_{\ \ \ \b}
+ R  \bar{\Gamma}^{\rho \omega\gamma}
- \eta^{\omega\gamma} R \bar{\Gamma}^{\rho \b}_{\ \ \ \b}] 
\\
+ 2  A  \partial_\a \partial_\omega [- R^{\a\gamma\rho\d}  h^\omega_{\ \d}   
- R^{\a\rho\gamma\d}  h^\omega_{\ \d}  
+ R^{\a\rho\omega\d}  h^\gamma_{\ \d}   
+ R^{\a\gamma\omega\d}  h^\rho_{\ \d}
+ R^{\a\omega\rho\d}  h^\gamma_{\ \d}  
+ R^{\a\omega\gamma\d}  h^\rho_{\ \d}]
\\
+ B \partial_\a \partial_\omega [ \eta^{\gamma\rho}   R^{\a\d}   h^\omega_{\ \d}  
-  \eta^{\rho\omega}   R^{\a\d}   h^\gamma_{\ \d}
-   \eta^{\gamma\omega}  R^{\a\d}   h^\rho_{\ \d}
+  R^{\a\omega}  h^{\gamma \rho}
+ R^{\gamma\rho}  h^{\omega \a} 
-  R^{\rho\omega}  h^{\gamma \a}
-  R^{\gamma\omega}  h^{\rho \a}
] 
\\
+  \frac{1}{2} B  \partial_\a \partial_\omega [ - \eta^{\a\rho}   R^{\gamma\d}  h^\omega_{\ \d}
- \eta^{\a\gamma}   R^{\rho\d}  h^\omega_{\ \d}
+  \eta^{\a\omega} R^{\rho\d}  h^\gamma_{\ \d} 
+  \eta^{\a\omega}  R^{\gamma\d}  h^\rho_{\ \d} 
 +   \eta^{\a\rho}   R^{\omega\d}  h^\gamma_{\ \d}
+  \eta^{\a\gamma}   R^{\omega\d}  h^\rho_{\ \d}] 
\\
+  2  C  \partial_\a \partial_\omega [\eta^{\a\omega} R  h^{\gamma \rho} 
+ \eta^{\gamma\rho} R  h^{\omega \a}
-     \eta^{\rho\omega} R  h^{\gamma \a} 
-     \eta^{\gamma\omega}  R  h^{\rho \a}] .
\end{multline}

Separating the part proportional to $ \eta^{\gamma \rho} $,

\begin{multline}
T^{\gamma\rho}_H = \eta^{\gamma \rho} \Bigg(
 A  R^{\a \b \c \d} R_{\a \b \c \d} + C  (R^2  + 8  \partial_\omega [ R \bar{\Gamma}^{\omega \b}_{\ \ \ \b} ]  +  2 \partial_\a \partial_\omega [  R  h^{\omega \a}] )
\\
+  B ( R^{ \b \d} R_{\b \d}  + 4 \partial_\omega [   R^{\b\d} \bar{\Gamma}^\omega_{\ \b \d}]  + \partial_\a \partial_\omega [  R^{\a\d}   h^\omega_{\ \d}   ] ) \Bigg)
\\
\\ 
-  8 A R^{\gamma \b \c \d} R_{\ \b \c \d}^{\rho} 
-  B  (4 R^{ \gamma \d} R_{\d}^{ \rho} + 4 R^{\gamma \b \rho \d} R_{\b \d} )
- 8 C R^{\gamma  \rho} R 
\\
\\
+ 16 A   \partial_\omega [R^{\gamma\b\rho\d} \bar{\Gamma}^\omega_{\ \b \d}
 - R^{\rho\b\omega\d} \bar{\Gamma}^\gamma_{\ \b \d}
 - R^{\gamma\b\omega\d} \bar{\Gamma}^\rho_{\ \b \d}]
\\
+ 4 B   \partial_\omega [- R^{\rho\d}  \bar{\Gamma}^{\omega \gamma}_{\ \ \ \d}
+ R^{\gamma\rho} \bar{\Gamma}^{\omega \b}_{\ \ \ \b}
- R^{\gamma\d}  \bar{\Gamma}^{\omega \rho}_{\ \ \ \d}
+ R^{\omega\d}  \bar{\Gamma}^{\gamma \rho}_{\ \ \ \d} 
 - \eta^{\omega\rho} R^{\b\d} \bar{\Gamma}^\gamma_{\ \b \d}] 
\\
 + 4 B  \partial_\omega [- R^{\rho\omega} \bar{\Gamma}^{\gamma \b}_{\ \ \ \b}
+ R^{\omega\d}  \bar{\Gamma}^{\rho \gamma}_{\ \ \ \d}
-  \eta^{\omega\gamma} R^{\b\d} \bar{\Gamma}^\rho_{\ \b \d}
-  R^{\gamma\omega} \bar{\Gamma}^{\rho \b}_{\ \ \ \b} 
+ R^{\rho\d}  \bar{\Gamma}^{\gamma \omega}_{\ \ \ \d}
+ R^{\gamma\d}  \bar{\Gamma}^{\rho \omega}_{\ \ \ \d}]  
\\
+ 8 C \partial_\omega [- R  \bar{\Gamma}^{\omega \rho\gamma}
+ R  \bar{\Gamma}^{\gamma \omega\rho}
- \eta^{\omega\rho} R \bar{\Gamma}^{\gamma \b}_{\ \ \ \b}
+ R  \bar{\Gamma}^{\rho \omega\gamma}
- \eta^{\omega\gamma} R \bar{\Gamma}^{\rho \b}_{\ \ \ \b}] 
\\
\\
+ 2  A  \partial_\a \partial_\omega [- R^{\a\gamma\rho\d}  h^\omega_{\ \d}   
- R^{\a\rho\gamma\d}  h^\omega_{\ \d}  
+ R^{\a\rho\omega\d}  h^\gamma_{\ \d}   
+ R^{\a\gamma\omega\d}  h^\rho_{\ \d}
+ R^{\a\omega\rho\d}  h^\gamma_{\ \d}  
+ R^{\a\omega\gamma\d}  h^\rho_{\ \d}]
\\
+ B \partial_\a \partial_\omega [ 
-  \eta^{\rho\omega}   R^{\a\d}   h^\gamma_{\ \d}
-   \eta^{\gamma\omega}  R^{\a\d}   h^\rho_{\ \d}
+  R^{\a\omega}  h^{\gamma \rho}
+ R^{\gamma\rho}  h^{\omega \a} 
-  R^{\rho\omega}  h^{\gamma \a}
-  R^{\gamma\omega}  h^{\rho \a}
] 
\\
+  \frac{1}{2} B  \partial_\a \partial_\omega [ - \eta^{\a\rho}   R^{\gamma\d}  h^\omega_{\ \d}
- \eta^{\a\gamma}   R^{\rho\d}  h^\omega_{\ \d}
+  \eta^{\a\omega} R^{\rho\d}  h^\gamma_{\ \d} 
+  \eta^{\a\omega}  R^{\gamma\d}  h^\rho_{\ \d} 
 +   \eta^{\a\rho}   R^{\omega\d}  h^\gamma_{\ \d}
+  \eta^{\a\gamma}   R^{\omega\d}  h^\rho_{\ \d}] 
\\
+  2  C  \partial_\a \partial_\omega [\eta^{\a\omega} R  h^{\gamma \rho} 
-     \eta^{\rho\omega} R  h^{\gamma \a} 
-     \eta^{\gamma\omega}  R  h^{\rho \a}] .
\end{multline}

It should be obvious at this point that there is no way we can reconcile this Hilbert energy-momentum tensor with what is uniquely derived from Noether's theorem for the linearized Gauss-Bonnet model, $T^{\omega\nu}_N $ in Equation (\ref{gaussemt}), if we fix coefficients $A = \frac{1}{4}$, $B = - 1$ and $C = \frac{1}{4}$ above. There is also a difference in corresponding coefficients between the terms the two share in common ($  \eta^{\gamma\rho} [A  R^{\a \b \c \d} R_{\a \b \c \d}+  B  R^{ \b \d} R_{\b \d}  + C  R^2]-  8 A R^{\gamma \b \c \d} R_{\ \b \c \d}^{\rho} -  B  (4 R^{ \gamma \d} R_{\d}^{ \rho} + 4 R^{\gamma \b \rho \d} R_{\b \d} ) - 8 C R^{\gamma  \rho} R $), meaning even this part is not equivalent. To prove that this does not equal to the Noether energy-momentum tensor ($T^{\omega\nu}_N \neq T^{\gamma\rho}_H$) we simply can compare the part proportional to $\eta^{\gamma \rho}$. The rest we will abbreviate as $\mathcal{T}^{\gamma\rho}_{HNMP}$ to represent the Hilbert non-Minkowski part. This allows us to write the Hilbert energy-momentum tensor in compact form,

\begin{multline}
T^{\gamma\rho}_H = \eta^{\gamma \rho} \Bigg(
 A  R^{\a \b \c \d} R_{\a \b \c \d} + C  (R^2  + 8  \partial_\omega [ R \bar{\Gamma}^{\omega \b}_{\ \ \ \b} ]  +  2 \partial_\a \partial_\omega [  R  h^{\omega \a}] )
\\
+  B ( R^{ \b \d} R_{\b \d}  + 4 \partial_\omega [   R^{\b\d} \bar{\Gamma}^\omega_{\ \b \d}]  + \partial_\a \partial_\omega [  R^{\a\d}   h^\omega_{\ \d}   ] ) \Bigg) + \mathcal{T}^{\gamma\rho}_{HNMP} .
\end{multline}

\subsection{Comparing the Noether and Hilbert energy-momentum tensors for linearized Gauss-Bonnet gravity}

We will now fix coefficients of the Hilbert expression for the specific counter-example of linearized Gauss-Bonnet gravity. Setting $A = \frac{1}{4}$, $B = - 1$ and $C = \frac{1}{4}$ yields,

\begin{multline}
T^{\gamma\rho}_H = \frac{1}{4} \eta^{\gamma \rho} \Bigg(
  R^{\a \b \c \d} R_{\a \b \c \d} - 4 R^{ \b \d} R_{\b \d}   +  R^2  
\\
+ 8  \partial_\omega [ R \bar{\Gamma}^{\omega \b}_{\ \ \ \b} ]  +  2 \partial_\a \partial_\omega [  R  h^{\omega \a}]  - 16 \partial_\omega [   R^{\b\d} \bar{\Gamma}^\omega_{\ \b \d}]  - 4 \partial_\a \partial_\omega [  R^{\a\d}   h^\omega_{\ \d}   ]  \Bigg) + \mathcal{T}^{\gamma\rho}_{HNMP} .
\end{multline}

Writing the Noether energy-momentum tensor in terms of the non-Minkowski part abbreviated as $\mathcal{T}^{\gamma\rho}_{NNMP} = -  R^{\omega\rho\lambda\sigma} R^\nu_{\ \rho\lambda\sigma} + 2  R_{\rho\sigma} R^{\omega \rho \nu \sigma} + 2  R^{\omega\lambda} R^\nu_{\ \lambda} -  R R^{\omega\nu}$ gives,

\begin{equation}
T^{\omega\nu}_N =  \frac{1}{4} \eta^{\omega\nu}(R_{\mu\lambda\alpha\beta} R^{\mu\lambda\alpha\beta} - 4 R_{\mu\gamma} R^{\mu\gamma} + R^2) + \mathcal{T}^{\gamma\rho}_{NNMP} .
\end{equation}

Subtracting the expressions for the Hilbert and Noether energy-momentum tensors given above yields,

\begin{equation}
T^{\gamma\rho}_H - T^{\omega\nu}_N = \frac{1}{4} \eta^{\gamma \rho} \Bigg(
8  \partial_\omega [ R \bar{\Gamma}^{\omega \b}_{\ \ \ \b} ]  +  2 \partial_\a \partial_\omega [  R  h^{\omega \a}]  - 16 \partial_\omega [   R^{\b\d} \bar{\Gamma}^\omega_{\ \b \d}]  - 4 \partial_\a \partial_\omega [  R^{\a\d}   h^\omega_{\ \d}   ]  \Bigg) + \mathcal{T}^{\gamma\rho}_{HNMP} - \mathcal{T}^{\gamma\rho}_{NNMP} .
\end{equation}

It is obvious $\mathcal{T}^{\gamma\rho}_{HNMP} - \mathcal{T}^{\gamma\rho}_{NNMP} \neq 0$. But to prove it for the $\eta^{\gamma \rho}$ part we will expand the derivatives,

\begin{multline}
T^{\gamma\rho}_H - T^{\omega\nu}_N = \frac{1}{4} \eta^{\gamma \rho} \Bigg(
(\eta^{\b\d} R  - 2 R^{\b\d}) (- 4 \partial_\omega \partial^\omega h_{\b\d} + 10 \partial_\omega  \partial_\d h^{\omega}_{\ \b}  )
\\
+  (\eta^{\b\d} \partial_\omega R - 2 \partial_\omega R^{\b\d}) (- 4 \partial^\omega h_{\b\d} + 10 \partial_\d h^{\omega}_{\ \b}   )   \Bigg) 
+ \mathcal{T}^{\gamma\rho}_{HNMP} - \mathcal{T}^{\gamma\rho}_{NNMP} .
\end{multline}

None of the remaining terms cancel. Therefore we have proven that $T^{\omega\nu}_N \neq T^{\gamma\rho}_H$ in the case of linearized Gauss-Bonnet gravity. This completes the disproof by counterexample of the notion that the Hilbert and Noether energy-momentum tensors are, in general, equivalent.

\section{Why the Noether and Hilbert energy-momentum tensors are not, in general, equivalent}

We have now disproved by counterexample the conjecture that the Noether and Hilbert energy-momentum tensors are, in general, equivalent. This notion is often asserted based on the fact that for simple models such as electrodynamics, with first order derivatives of a vector potential in the Lagrangian, the two methods indeed yield the same result. The best attempted proofs of equivalence have relied on assuming first order derivatives in the action \cite{forger2004,pons2011}, as discussed in the Motivation section. Therefore we considered a model with second order derivatives of a second rank tensor potential in the Lagrangian (linearized higher order gravity models), and in particular the linearized Gauss-Bonnet gravity model which has a unique and well established energy-momentum tensor. In this case, the Noether and Hilbert energy-momentum tensors are not equivalent.

In this section we will conclude by explaining the three major reasons why
the Noether and Hilbert methods for deriving an energy-momentum tensor in Minkowski spacetime diverge from
one another for models with higher order derivatives and higher rank tensor potentials.

{\bf{Reason 1:}} The Noether energy-momentum tensor $T^{\omega\nu}_N$ has terms proportional to  $\eta^{\gamma\omega}$ which are exactly the Lagrangian density $\eta^{\gamma\omega} \mathcal{L}$ as seen in equation (\ref{genenergy}). This piece is always exactly proportional to the        Lagrangian, which follows directly from application of Noether's first theorem. The Hilbert method, for models such as the linearized gravity models defined in equation (\ref{alllingravlags}), produces terms proportional to $\eta^{\gamma\omega}$ beyond what is present in the Lagrangian. In such cases, the Noether and Hilbert methods yield different results.

{\bf{Reason 2:}} The covariant derivatives of higher rank tensor potentials do not cancel as in the case of simple models like the scalar field (Klein-Gordon) and vector field (electrodynamics). For the covariant derivative of a scalar we have only $\nabla_\mu \phi = \partial_\mu \phi$, therefore we get no part with Christoffel symbols in the curved space Lagrangian. In electrodynamics, again we have $F^\nabla_{\mu\nu} = \nabla_\mu A_\nu - \nabla_\nu A_\mu = \partial_\mu A_\nu - \Gamma^\alpha_{\mu\nu} A_\alpha - \partial_\nu A_\mu + \Gamma^\alpha_{\nu\mu} A_\alpha = F_{\mu\nu}$, where the $\Gamma^\alpha_{\mu\nu}$ contributions exactly cancel due to the antisymmetry of the field strength tensor, thus there is again no Christoffel part to the curved space Lagrangian. For models where the Christoffel symbols do not cancel, such as the linearized higher derivative gravity models given in equation (\ref{alllingravlags}) and considered in this paper, there are many lingering contributions from Christoffel symbols that simply add many terms which do not follow from Noether's method.

{\bf{Reason 3:}} The proportional to Minkowski piece ($\eta^{\gamma\omega} \mathcal{L}$) derived from Noether's theorem in Equation (\ref{genenergy}) is recovered from the Hilbert method by differentiating the $\sqrt{-g}$ part of the curved space action. The remaining terms are built by contracting the expression with the metric. In electrodynamics this scalar is built from 2 metrics, for the linearized gravity model we consider there will be 4 metrics, and so on. The higher the rank of tensors one builds a Lagrangian from, the more metric contractions the curved space Lagrangian will have, but there will always be only one $\sqrt{-g}$ contribution. Therefore the relative contribution of the $\eta^{\gamma\omega}$ piece to the non $\eta^{\gamma\omega}$ piece in the Hilbert expression differs as we increase the order of derivatives and rank of fields, yet the  $\eta^{\gamma\omega} \mathcal{L}$ piece from the Noether method has the same relationship to the non $\eta^{\gamma\omega}$ part regardless of the rank of tensor or order of derivatives. In simple cases such as a first order scalar field (Klein-Gordon), and first order vector field (electrodynamics) the proportion of these two contributions in the Hilbert method is identical to the Noether method. What this means is that even if we ignore problematic Christoffel terms from Reason 3, the Hilbert energy-momentum tensor would still not coincide with the Noether energy-momentum tensor.

These reasons provide some intuitive insight into why the Noether and Hilbert energy-momentum tensors are not in general giving the same result, without need to consider the details of the more technical disproof provided in this article. The obvious question that now arises is, if we have two methods for deriving energy-momentum tensors from a Lagrangian density of a model in Minkowski spacetime, and they are not generally giving equivalent results, which one should be considered like the fundamental method for deriving physical expressions? In the case of equations of motion derived from a Lagrangian density, the Euler-Lagrange equation has no such `different' method to ` compete' with. The connection of the Noether method to the Euler-Lagrange equation, coupled with its connection to symmetry and to the derivation of the unique and well accepted expression for linearized Gauss-Bonnet gravity used in the disproof by counterexample in this article, seem to speak for itself. Any more general, philosophical thoughts related to this decision will be left to the reader.

\section{Acknowledgement}

The authors are grateful to D.G.C. McKeon for numerous discussions and suggestions during the preparation of our article.

\section{Bibliography}

\bibliographystyle{elsarticle-num}
\bibliography{NoetherHilbertCounterexample}

\end{document}